\documentclass[aps,prx,twocolumn,superscriptaddress,nofootinbib,floatfix]{revtex4-1}
\usepackage{amsmath,amssymb,wasysym,graphicx}
\usepackage{times}
\usepackage[varg]{txfonts}
\usepackage{textcomp}
\usepackage{tabularx}
\usepackage{subfigure}
\usepackage{tabu}
\usepackage{color}
\usepackage[colorlinks=true,citecolor=blue,urlcolor=blue,linkcolor=blue,hyperindex]{hyperref}
\usepackage{braket}
\usepackage{overpic}
\usepackage{appendix}

\usepackage[normalem]{ulem}
\usepackage{verbatim}
\allowdisplaybreaks

\begin{document}
\title{Detecting Subsystem Symmetry Protected Topological Order  Through Strange Correlators}
 \author{Chengkang Zhou}
\thanks{These authors contributed equally to this work.}
\affiliation{Department of Physics and HKU-UCAS Joint Institute of Theoretical and Computational Physics,The University of Hong Kong, Pokfulam Road, Hong Kong SAR, China}
\author{Meng-Yuan Li}
\thanks{These authors contributed equally to this work.}
\affiliation{School of Physics, State Key Laboratory of Optoelectronic Materials and Technologies, and Guangdong Provincial Key Laboratory of Magnetoelectric Physics and Devices, Sun Yat-sen University, Guangzhou, 510275, China}
\author{Zheng Yan}
\affiliation{Department of Physics and HKU-UCAS Joint Institute of Theoretical and Computational Physics,The University of Hong Kong, Pokfulam Road, Hong Kong SAR, China}
\author{Peng Ye}\email{yepeng5@mail.sysu.edu.cn}
\affiliation{School of Physics, State Key Laboratory of Optoelectronic Materials and Technologies, and Guangdong Provincial Key Laboratory of Magnetoelectric Physics and Devices, Sun Yat-sen University, Guangzhou, 510275, China}
\author{Zi Yang Meng}\email{zymeng@hku.hk}
\affiliation{Department of Physics and HKU-UCAS Joint Institute of Theoretical and Computational Physics,The University of Hong Kong, Pokfulam Road, Hong Kong SAR, China}

\begin{abstract}
We   employ \emph{strange correlators}  to detect   2D subsystem symmetry protected topological (SSPT) phases which are nontrivial topological phases protected by subsystem symmetries. Specifically, we    analytically construct   efficient strange correlators in the 2D cluster model in the presence of uniform magnetic field, and then  perform the projector Quantum Monte Carlo   simulation within the quantum annealing scheme.  We find that strange correlators show long range correlation in the SSPT phase, from which we define \emph{strange order parameters} to characterize  the topological phase transition between the SSPT phase at low fields and  the trivial paramagnetic phase at high fields.  Thus, the detection of the fully localized zero modes on the 1D physical boundary of SSPT phase has been transformed to the bulk correlation measurement about the local operators with the periodic boundary condition.   We also find interesting spatial anisotropy of a strange correlator, which can be intrinsically traced back to  the nature of spatial anisotropy of subsystem symmetries that protect SSPT order in the 2D cluster model. By simulating strange correlators, we therefore  provide the first unbiased large-scale quantum Monte Carlo simulation on the easy and efficient detection in the SSPT phase and open the avenue of the investigation of the subtle yet fundamental nature of the novel interacting topological phases.
\end{abstract}
\date{\today}
	
\maketitle
\section{Introduction}
\label{sec:Introduction}

Exemplified by the Haldane chain, bosonic symmetry protected topological (SPT) phases in interacting boson/spin systems are short-range entangled states protected by  global symmetry~\cite{Chen10,Chen_science,PhysRevB.81.064439,1DSPT,Chen2013CGLWbSPT}. SPT phases are sharply distinct from   topological orders~\cite{RevModPhys.89.041004,2DbTO_wen_15} that admit   long-range entanglement and robustly persist  regardless of symmetry protection. More concretely, SPT phases  can be adiabatically connected to trivial direct product states by local unitary transformations if symmetry is allowed to be broken. However, such adiabatic paths are forbidden if symmetries are respected. For the past decades, SPT phases have been intensively studied through different approaches including  group cohomology~\cite{Chen2013CGLWbSPT}, cobordism groups~\cite{Kapustin2014Cobordism, Kapustin2015fCobordism}, non-linear sigma models~\cite{Bi2015NLSMSPT,YouYou2016bSPT}, topological field theories ~\cite{Lu2012CSSPT,bti2,Gu2016SPTBF,Ye20163dbSPT,2018arXiv180105416W}, conformal field theories~\cite{Hsieh2014SPT2dorbifold,Hsieh2016SPT3dorbifold,Han2017BCFTSPT}, decoration picture~\cite{Chen:2014aa}, topological response/gauged theory ~\cite{PhysRevLett.112.141602,Hung_Wen_gauge,PhysRevD.88.045013,bti6,lapa17,Wang2015GaugeGravitySPT,PhysRevB.99.205120},  projective/parton construction~\cite{YW12,LL1263,Ye14b,YW13a,ye16a,Wangsenthil2015}, braiding statistics approach~\cite{levin_gu_12,wang_levin1,2016arXiv161209298P,ye17b}, and strange correlators ~\cite{Wave2014You,Quantum2015Wu,Strange2014Wierschem,Detection2016Wierschem,Bona2016He,Topological2017Zhong}, which ignites great research interests and joint efforts from condensed matter physics, mathematical physics, and quantum information.

However, as short-range entangled states, SPT phases are non-fractionalized in the bulk and are not characterized by the more interesting properties found in  topological orders. For example, the behavior of entanglement entropy of SPT phases is trivially dominated by area law, which is a common feature for all gapped phases. In contrast, topological orders admit exotic sub-leading term---topological entanglement entropy---which is quantitatively determined by total quantum dimension  of  anyons~\cite{Kitaev2006,Levin2006,zhaoMeasuring2022,chenTopological2022}. Also in quantum critical points and other gapless systems, the entanglement entropy admits logarithmic corrections with coefficient representing the conformal field theory content~\cite{Fradkin2006,Casini2006,zhaoScaling2022,yaoRelating2021}. In all these situations, the strong entanglement becomes the organizing principle of highly entangled quantum matter. Therefore, one of   subsequent directions, in the field of SPT physics, is to find ways to   design and detect new types of SPT phases with potentially richer entanglement properties despite the absence of fractionalization in the bulk.

Along this line of thinking and motivated by the field of fracton physics~\cite{Nandkishore2019,2020Fracton,Chamon05fracton, Haah11, Yoshida13, Fu16,Nandkishore2019,2020Fracton,Hermele17, Vijay17coupledLayersXcube, Shirley2018, Chen20coupledCS, Aasen20tqftnetwork, Slagle21foliatedfield, LiYe20fracton, LiYe21fracton,XuWu08RVP, Pretko17tensorgauge, Chen18tensorhiggs, Barkeshli18tensorhiggs, Seiberg21ZNfracton,Pretko18fractongaugeprinciple, Gromov19multipoleAlgebra, Seiberg19vectorSym,2020PhRvR2b3267Y,Chen2021FS,Ye2021PRR,Yuan:2022mns,Zhu:2022lbx,Nandkishore21sym,PhysRevD.104.105001,Bidussi:2021nmp,Jain:2021ibh,Angus:2021jvm,Grosvenor:2021hkn,Banerjee:2022unj,Nandkishore21sym,Pretko2018,PhysRevX.9.031035,Seiberg2019arXiv1909,Gorantla:2022eem}, recently, an exotic class of SPT orders dubbed as subsystem symmetry protected topological (SSPT) orders was proposed~\cite{Subsystem2018You,Classification2018Devakul,Anomaly2021Burnell,Bifurcating2021SanMiguel,Entanglement2019Schmitz,Interaction2022MayMann,Subsystem2018Stephen,Symmetric2020You,Universal2018Devakul}, and shows a series of intriguing properties beyond aforementioned SPT phases protected by global symmetries, such as spurious topological entanglement entropy~\cite{Spurious2019Williamson,Spurious2016Zou,Toy2020Kato,Detecting2019Stephen} and duality into fracton topological orders~\cite{Subsystem2018You,Subsystem2018Stephen,Foliated2019Shirley,Entanglement2019Schmitz,Symmetric2020You,Strong2020Devakul,Anomaly2021Burnell,Bifurcating2021SanMiguel,Interaction2022MayMann}. Besides, SPT protected by fractal subsystem symmetries~\cite{Fractal2019Devakul, Universal2018Devakul, Classifying2019Devakul}, subsystem symmetry enriched topological orders~\cite{Subsystem2020Stephen}, higher order topological superconductors protected by subsystem symmetries~\cite{Higher2019You} and the computational properties of SSPT phases~\cite{Ungauging2018Kubica, Computational2020Daniel, Universal2018Devakul} have also been discussed. The subsystem symmetry refers to a kind of symmetries that by definition lies between global and local (gauge) symmetries~\cite{Discrete2005Nussinov,Generalized2005Batista,Generalized2022McGreevy}, which is, in this conceptual level, similar to the higher rank symmetry~\cite{2020PhRvR2b3267Y,Chen2021FS,Ye2021PRR,Yuan:2022mns} despite the subtle difference. In other words, a local symmetry can act on degrees of freedom inside an area that is negligible at the thermodynamic limit, while a global symmetry acts on degrees of freedom from all of the system. As an interpolation between global and gauge symmetry transformations, a subsystem symmetry transformation acts on degrees of freedom inside a subdimensional area which is sub-extensive at thermodynamic limit. Thus, the definition of  SSPT phases is clear: nontrivial SPT phase protected by a subsystem symmetry that is generated by subextensively infinite number of symmetry generators. For example, in Ref.~\cite{Subsystem2018You}, the 2D cluster model~\cite{Persistent2001Briegel}, which is also denoted as topological plaquette Ising model (TPIM), is identified as having an SSPT order protected by  a linear subsystem symmetry. By \emph{linear}, we mean that the Hamiltonian is invariant under flipping  spins sitting along a straight line, i.e., a subsystem of the whole 2D spin model.
The number of such straight lines is subextensively infinite.

Theoretically, Ref.~\cite{Subsystem2018You} finds that  the 1D edge states protected by such unconventional  symmetries are fully localized zero modes that  are exactly on   the endpoints of symmetry generators (see Sec.~\ref{sec:Model}). Therefore, such highly localized (dispersionless) edge modes of SSPT phases actually cast shadow in the SSPT detection and render  featureless results in traditional correlation-based theoretical analysis and numerical simulation in transport and spectroscopic  measurements, compared with the easy detection with such means of their 2D SPT cousins protected by global symmetries. Furthermore, creating a clean boundary with less finite-size effect is challenging both in experiments and large-scale simulations. Especially, in a practical quantum Monte Carlo (QMC) simulation, it is generally    convenient to compute   correlation function-like observables, but unfortunately all gapped topological phases including SPT, SSPT and topological orders do not show any defining properties in such easily accessible observables in the bulks.

To apply QMC simulation on SSPT order with a periodic boundary condition, in this paper we  attempt to take full advantage of  \emph{strange correlators} initially proposed in Ref.~\cite{Wave2014You}.  While more details about strange correlators are present in the main text, here let us briefly introduce the notion of strange correlators.  The ``strange'' definition of such correlators is that the bra- and ket- wavefunctions are respectively a trivial symmetric direct-product state and the state to be diagnosed. Then, strange correlators are defined as such a strange type of two-point correlation of a local operator $\phi$. In contrast to traditional correlation functions that exponentially decay in topological phases without symmetry-breaking orders, strange correlators of some $\phi$ will either saturate to a constant or show a power law decay at long distances for a non-trivial SPT state. Since the calculation of strange correlators are performed with periodic boundary condition, one obvious advantage of strange correlators is that a spatial interface (i.e., physical boundary) between the trivial state and the ground state to be diagnosed is unnecessary, rendering  bulk measurement of correlation functions of strange type.  In the literature,  a series of SPT orders protected by global symmetries, including free and interacting fermion topological insulators, the Haldene phase and some other exotic SPT phases, have been successfully detected by strange correlators~\cite{Quantum2015Wu,Strange2014Wierschem,Wave2014You,Detection2016Wierschem,Bona2016He,Topological2017Zhong}. Besides, the theoretical idea of strange correlator has also been utilized and generalized in the study of various topics, including intrinsic topological orders and conformal field theories\cite{Wave2016Scaffidi,Field2016Takayoshi,Global2018Bultinck,Gapless2017Scaffidi,Mapping2018Vanhove,Cardy2019Lootens,Galois2020Lootens,holographic2020McMahon,Self2021Fan,Universal2021Wu,Characterizing2022Lu,Construction2022Noh,Critical2022Vanhove,Topological2022Vanhove,Strange2022Lepori}. While there has been exciting progress on numerical simulation of fracton ordered lattice models as well as systems with subsystem symmetries~\cite{Evolution2022zhou,Correlation2018Devakul,Detecting2019Stephen,Entanglement2017Helmes,Topological2022Zhu,Quantum2020Muhlhauser,Anomalous2019Iaconis},  such a strange correlator diagnosis of SSPT phases is still lacking.

In this paper,  we  investigate  strange correlators of the 2D cluster model~\cite{Persistent2001Briegel,Subsystem2018You} via the projector QMC  method within the quantum annealing (QA) scheme where the SSPT-ordered ground state is accessible by sampling an operator strings acting on a trial state~\cite{Ground2005Sandvik,Loop2010Sandvik,Computational2010Sandvik}. We analytically study strange correlators of various local operators $\phi$ when exact solvability can be achieved. Based on the analytic results, we propose a set of SSPT-order-detectable  and  QMC-accessible strange correlators for the purpose of the large-scale numerical simulation on the effect of uniform magnetic fields that drive the system away from the exactly solvable point. Numerically, we find that in the presence of symmetry-respecting magnetic field, strange correlators show long range correlation in a finite range of magnetic field, and most importantly,   strange correlators unambiguously  signal the first-order phase transition between the SSPT phase and  the trivial paramagnetic phase. We also introduce the notion of  \emph{strange order parameter}, which is defined as the remaining finite value of strange correlators at long-distances (i.e., half of the linear size of the 2D system with the periodic boundary condition), in order to signal the existence of the SSPT phase by following  the general wisdom of Landau's theory of symmetry-breaking phases. In summary, by means of strange correlators,  the detection of  the fully localized zero modes on the 1D physical boundary  of SSPT phase has been transformed to the bulk correlation measurement about the local operators with  the periodic boundary condition. We also find interesting spatial anisotropy of a strange correlator, which can be intrinsically traced back to  the nature of spatial anisotropy of subsystem symmetries that protect SSPT order in the 2D cluster model. Our findings therefore provide the first unbiased large-scale quantum Monte Carlo simulation on the easy and efficient detection in the SSPT phase and open the avenue of the investigation of the subtle yet fundamental nature of the novel interacting topological phases.
Along with previous studies in conventional SPT physics~\cite{Quantum2015Wu,Strange2014Wierschem,Wave2014You,Detection2016Wierschem,Bona2016He,Topological2017Zhong}, the present study on SSPT phases provides new evidence of the effectiveness of strange correlators in characterizing nontrivial orders in topological phases of matter, and is expected to  stimulate the theoretical effort towards a  more systematic understanding  on strange correlators.

The rest of the paper is organized as follows. In Sec.~\ref{sec:Model}, the 2D cluster model for SSPT phase is introduced, focusing on the subsystem symmetries that protect the topological nature of the SSPT phase. In Sec.~\ref{sec:phase_diag}, we introduce the field-induced phase diagram of 2D cluster model. Then in Sec.~\ref{sec:StrangCorr}, the construction of the strange correlator in the present model is explained, along with the basic description of the projector QMC method employed (the detailed description of the projector QMC methodology is in   Appendix~\ref{sec:numerical method}). Sec.~\ref{sec:results} contains the main results of our work, where the strange correlator in strong and weak SSPT phases, and across their SSPT to trivial phase transition points by means of the transverse magnetic field, are presented and discussed. Based on these results, the advantageous usage of the strange correlator in the SSPT order detection is unambiguously shown. A few immediate future directions and conclusions are given in the Sec.~\ref{sec:discussion}.

\section{The 2D cluster model in the presence of Zeeman fields}
\label{sec:2DCModel}

\subsection{Model Hamiltonian and its sign-problem-free basis}
\label{sec:Model}
\begin{figure}[htp!]
	\centering
	\includegraphics[width=\columnwidth]{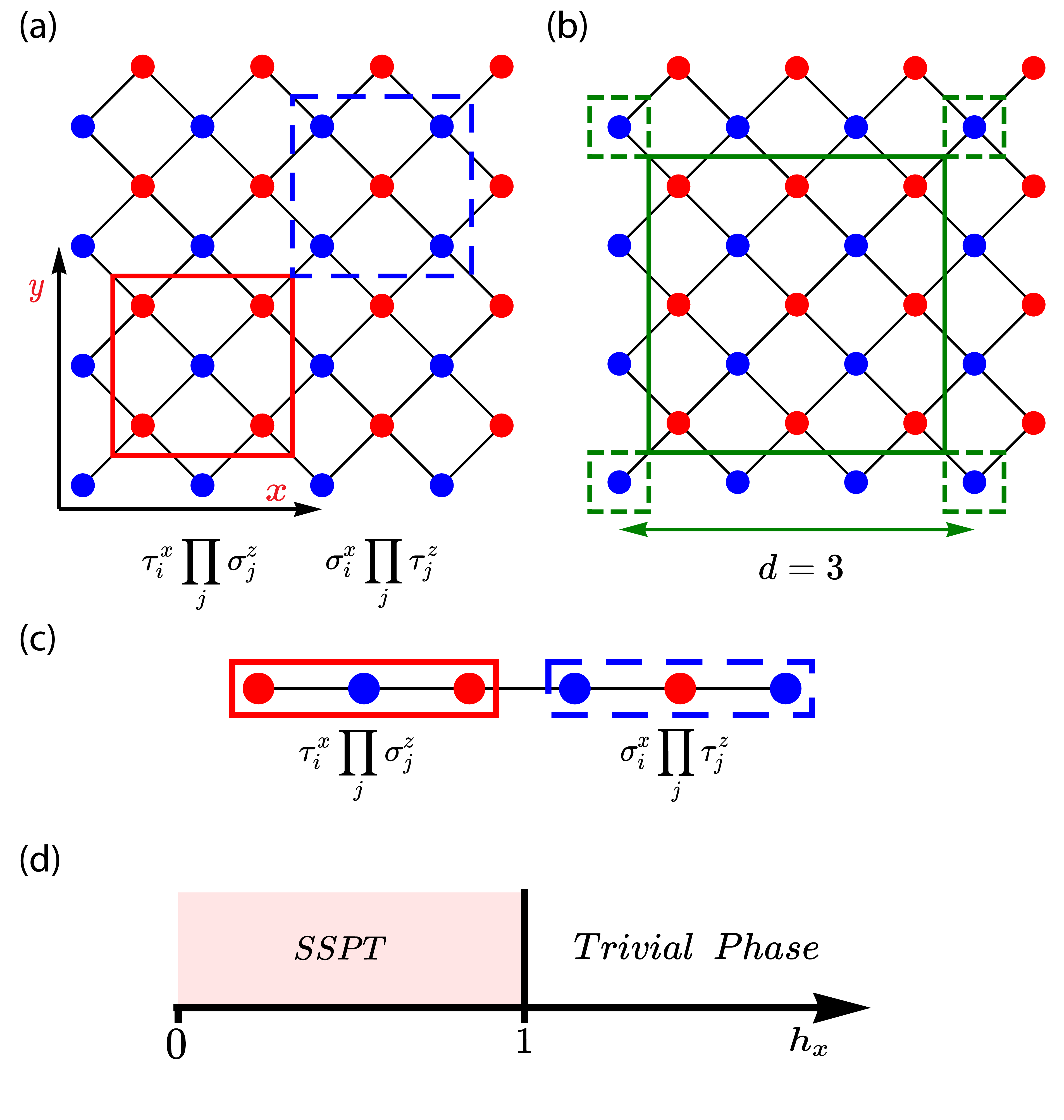}
	\caption{An illustration of the 2D and 1D cluster models. The blue points denote the $\tau$ spins and the red points denote the $\sigma$ spins. (a) describes the 2D cluster model with the red (blue) square telling the $A$ ($B$) term in Eq.~\eqref{eq:eq2}. (b) is the illustration of the membrane order parameter in Eq.~\eqref{eq:eq3} with the membrane size $d=3$. (c) is the illustration of the 1D cluster model and (d) is the phase diagram of 2D cluster model under the transverse field $h_{x}$.}
	\label{fig:fig_model}
\end{figure}
The Hamiltonian of the 2D cluster model defined on a square lattice is given as
\begin{equation}
	\begin{aligned}
		H&=-K\sum_{i}X_{i}\prod_{j}Z_{j}\,,
		\label{eq:eq1}
	\end{aligned}
\end{equation}
where $K>0$, $X_{i}$ and $Z_{i}$ are the Pauli matrices of the spin-$1/2$ degree of freedom living on site $i$ in the square lattice and $j$ for the four spin around the site $i$. Since $X_{i}$ is an off-diagonal operator in the $\{Z_{i}\}$ basis, the direct projector QMC simulation of the Hamiltonian Eq.~\eqref{eq:eq1} would meet the sign problem. In order to avoid this sign problem, we divide the square lattice into two sublattices and rewrite their associated spins respectively as $\tau^{x,y,z}_{i}$ and $\sigma^{x,y,z}_{i}$. And the index $i$ denotes the site of the unit cell, which contains one $\tau$ spin and one $\sigma$ spin. Under the external Zeeman magnetic fields, the 2D cluster model can be expressed as \cite{Classification2018Devakul,Subsystem2018You}
\begin{equation}
	\begin{aligned}
		H&=-K\sum_{i}A_{i}-K\sum_{i}B_{i}-h_{x}\sum_{i}\tau^{x}_{i}-h_{x}\sum_{i}\sigma^{x}_{i},
		\label{eq:eq2}
	\end{aligned}
\end{equation}
where $A_i\equiv \tau^{x}_{i}\prod_{j}\sigma^{z}_{j}$ and $B_i\equiv \sigma^{x}_{i}\prod_{j} \tau^{z}_{j}$. And we take $K=1$ as a unit in the following discussion. As shown in Fig.~\ref{fig:fig_model}(a), the red square refers to an $A$ term in Eq.~\eqref{eq:eq2} and the blue square is a $B$ term. Under this expression, taking the $\{\tau^{z}_{i}\sigma^{x}_{i}\}$ basis, $A$ terms are purely off-diagonal and $B$ terms are purely diagonal. Thus the sign problem can be avoided.

As shown in Ref.~\cite{Subsystem2018You}, the 2D cluster model is invariant under linear $\mathbb{Z}^{sub}_2$ subsystem symmetry transformations generated by $\prod_{i\in l_{k}} \sigma_{i}^x$ and $\prod_{i\in l_{k}} \tau_{i}^x$ operators, where $k=x,y$, $l_{x}$ ($l_y$) is an arbitrary straight line parallel to $x$- ($y$-) axis.
If we consider the edges of the 2D cluster model in Fig.~\ref{fig:fig_model} with open boundary condition, taking the upper edge for instance, for a truncated cluster with a $\sigma$ spin sitting on its center, we can define a set of operators: $\pi_{i}^x=\tau_{i}^z\sigma_{i}^x\tau_{i+\hat{x}}^z,\pi_{i}^y=\tau_{i}^z\sigma_{i}^y\tau_{i+\hat{x}}^z,\pi_{i}^z=\sigma_{i}^z$, where we set a unit cell to be composed of a $\tau$ spin located at $(a,b)$ and a $\sigma$ spin located at $(a+\frac{1}{2}\hat{x},b+\frac{1}{2}\hat{y})$, $\hat{x}$ and $\hat{y}$ are respectively the unit vectors along $x$- and $y$-directions (for truncated clusters with $\tau$ spins sitting on the center, the corresponding $\pi$ operators can be obtained by simply switching $\sigma$ and $\tau$). As these operators form an $SU(2)$ Lie algebra and simultaneously commute with all Hamiltonian terms, we can draw a conclusion that such a set of $\pi$ operators exactly form a two-dimensional  Hilbert space of the 2D cluster model, which is nothing but a free $\frac{1}{2}$-spin degree of freedom  localized on the site $i$. Thus, for each site on the upper edge, there is a localized dangling $1/2$ spin whose excitation energy is zero.  Such edge modes are protected by the corresponding linear subsystem symmetries. Therefore, unlike 2D SPT orders protected by global symmetries, in the 2D cluster model the degeneracy introduced by open boundary condition grows exponentially with the length of edge. Besides, as demonstrated in Ref.~\cite{Subsystem2018You}, the effective edge Hamiltonian cannot have any nonidentity local terms that respect all symmetries. That is to say, the protected edge modes are always nondispersing in the SSPT phase. Since the model at $h_x=0$ is exactly solvable, the ground state can also be analytically obtained, which is reviewed in Appendix~\ref{appendix_gs}.

\subsection{Field-induced phase diagram from projector QMC}
\label{sec:phase_diag}

In the presence of the transverse fields $h_{x}$, by duality transforming to the Xu-Moore model, a first order phase transition has been discovered at $h_{x,c}=1$, from the SSPT phase to a trivial paramagnetic phase (see Fig.~\ref{fig:fig_model}(d))~\cite{Strong2004Xu,Reduction2005Xu,First2009Orus,Fate2012Kalis,Subsystem2018You}. In order to characterize this phase transition, we compute three different physical observables, including the energy density $e$, the magnetization of spin $\sigma$ defined as $m_{\sigma^{x}}=\sum_{i}\sigma^{x}_{i}/N_{s}$ with $N_{s}$ referring to the total number of $\sigma$ spin, and the membrane order parameter $O_{d}$ defined as \cite{Identifying2009Doherty,Spurious2019Williamson}
\begin{equation}
	\begin{aligned}
		O_{d}&=\prod_{i\in C}\tau^{z}_{i}\prod_{j\in M}\sigma^{x}_{j}.
		\label{eq:eq3}
	\end{aligned}
\end{equation}
Here, $C$ refers to the corners of the membrane $M$, which corresponds to the blue spins ($\tau$) living inside the small dashed squares in Fig.~\ref{fig:fig_model} (b). And $M$ is a membrane, that is the collection of the red spins ($\sigma$) inside the solid squares in Fig.~\ref{fig:fig_model} (b). And the factor $d$ is the linear size of the membrane, which is shown by the green line for $d=3$ in Fig.~\ref{fig:fig_model} (b). In the SSPT phase, the values of $\langle O_{d} \rangle$ would approach a constant as $d\rightarrow\infty$, while it would tend to zero in the trivial paramagnetic phase.

\begin{figure}[htp!]
	\centering
	\includegraphics[width=\columnwidth]{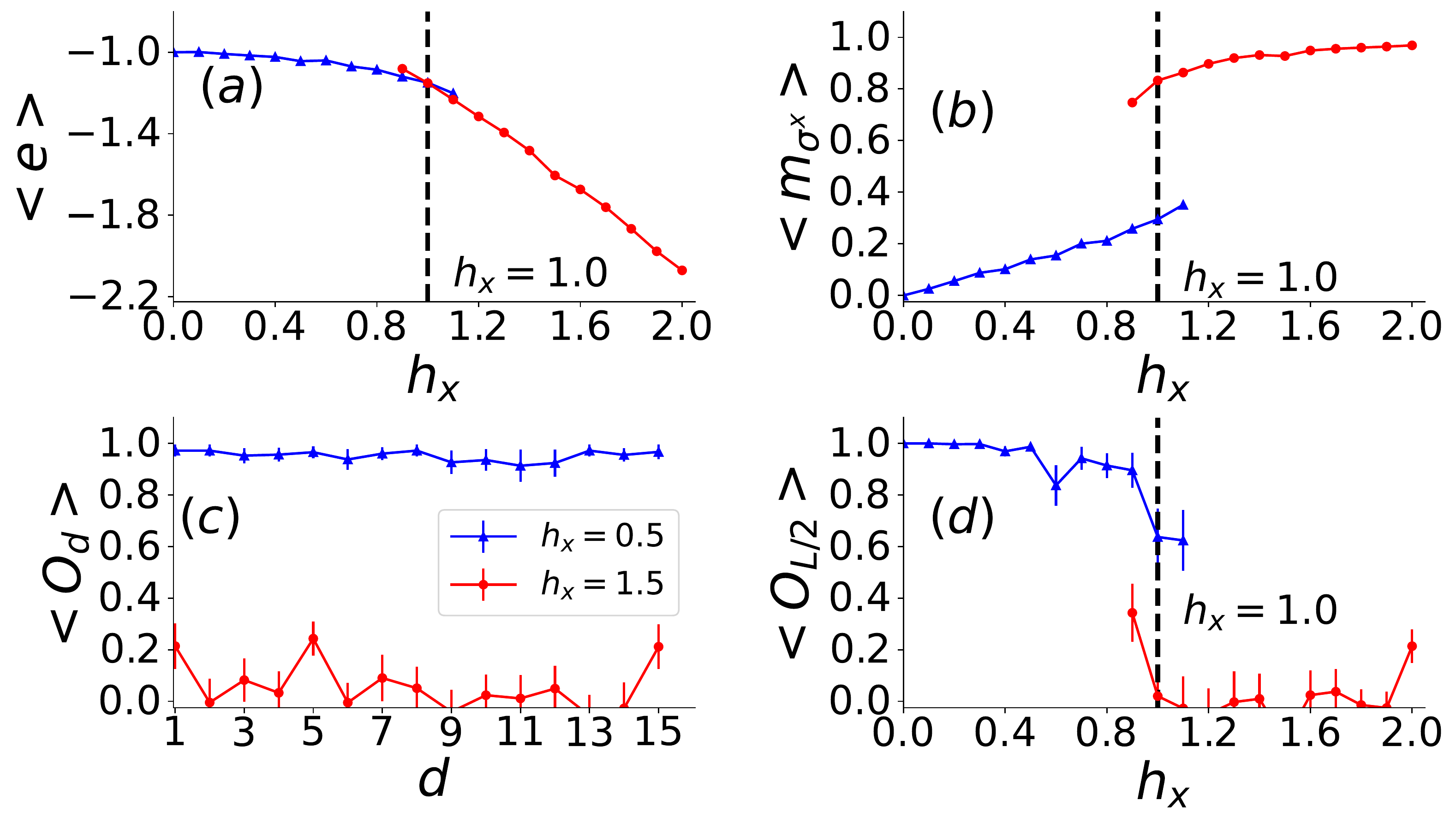}
	\caption{The values of different physical observables under the transverse field $h_{x}$. (a,b,d) are the energy per site, the magnetization of spin $\sigma$ along $x$-direction, and the membrane order parameter, respectively. Among them, the blue line with triangle are annealing from the exactly solvable point $h_{x}=0$ while the red line with circle is from the high field limit $h_{x}=2$. And (c) denotes the behavior of the membrane order parameter with increasing membrane size, in which the blue line is inside the SSPT phase ($h_{x}=0.5$) and the red is in the paramagnetic phase ($h_{x}=1.5$).}
	\label{fig:fig_normal_result_2D_hx}
\end{figure}

Using the projector QMC method and QA process (see details in Appendix~\ref{sec:appA1}), we have simulated the 2D cluster model with system size $L=16$ and annealing step $\Delta h=0.01$, and measured the energy density $\langle e \rangle$, the magnetization $\langle m_{\sigma^{x}} \rangle$ , and the membrane order parameter $\langle O_{d} \rangle$, which is shown in Fig.~\ref{fig:fig_normal_result_2D_hx}. Firstly, we observe $\langle O_{d} \rangle$ as a function of $d$, which is plotted as the blue line for the SSPT phase ($h_{x}=0.5$) and the red line for the trivial paramagnetic phase ($h_{x}=1.5$) in Fig.~\ref{fig:fig_normal_result_2D_hx} (c). As expected, $\langle O_{d} \rangle$ approaches a constant as the membrane size of $\langle O_{d} \rangle$ increasing in the SSPT phase. In contrast, $\langle O_{d} \rangle$ is zero for large membrane size in the trivial paramagnetic phase.

Meanwhile, to pin down the phase transition point of the 2D cluster model induced by the transverse field $h_{x}$, we have plotted the value of $\langle e \rangle$, $\langle m_{\sigma} \rangle$ , and $\langle O_{L/2} \rangle$ as a function of $h_{x}$ respectively in Figs.~\ref{fig:fig_normal_result_2D_hx} (a), (b) and (d). The blue lines in these figures with the triangle points are measured by scanning from the exactly solvable point ($h_{x}=0$) while the red lines with dot points are from the strong field limit ($h_{x}=2$). From Fig.~\ref{fig:fig_normal_result_2D_hx} (a), (b) and (d), a clear first order phase transition has been observed at $h_{x,c}=1$, which is consistent with the result of analytical mapping mentioned above.

\begin{figure}[htp!]
	\centering
	\includegraphics[width=\columnwidth]{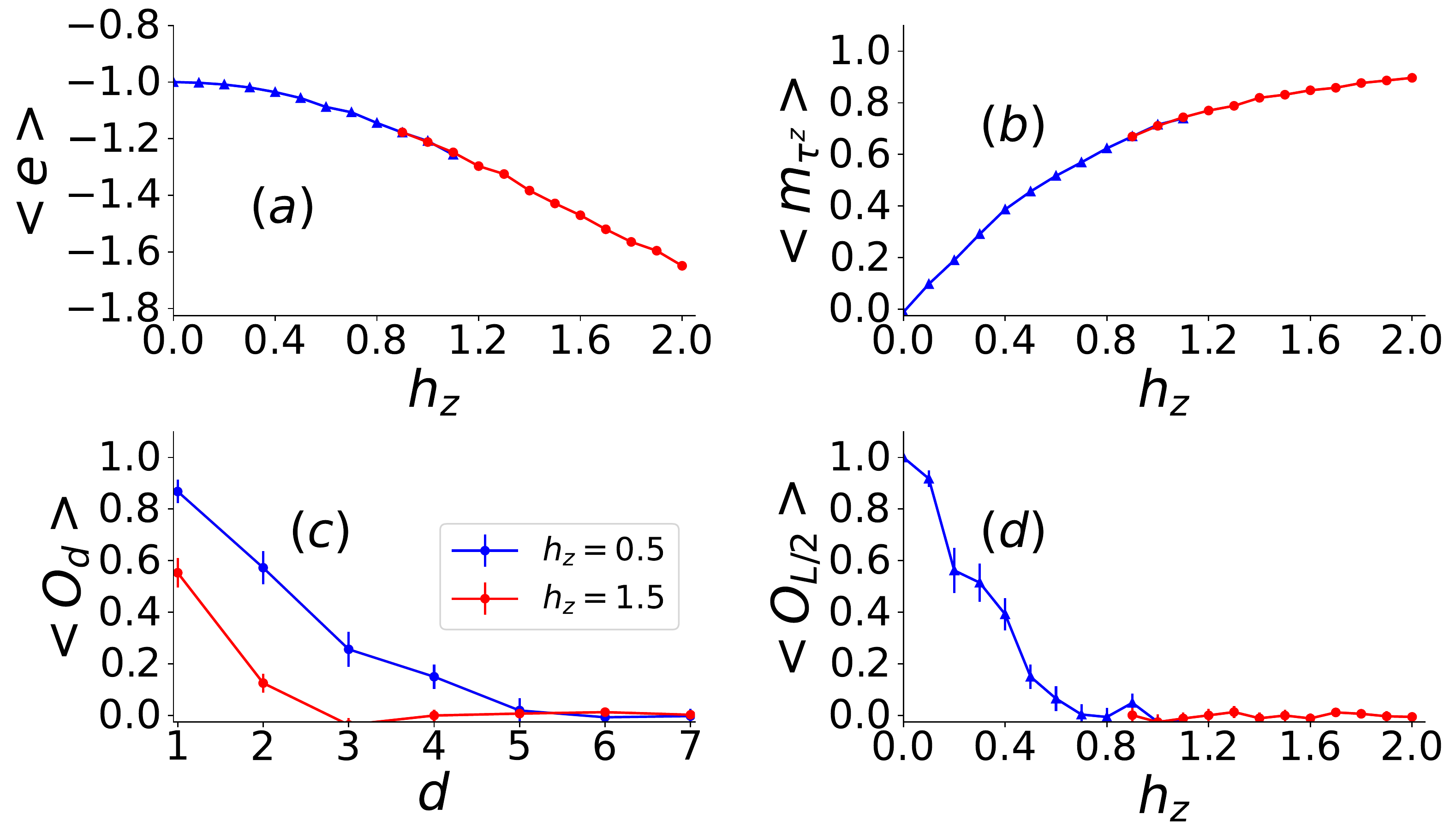}
	\caption{The values of different physical observables under the longitudinal field $h_{z}$. (a) and (b) are respectively the energy per site and magnetization of spin $\tau$ along $z$-direction. Among them, the blue lines with triangles are annealing from the exactly solvable point $h_{z}=0$ while the red lines with dots is from the strong field limit $h_{z}=2$.}
	\label{fig:fig_normal_result_2D_hz}
\end{figure}

In addition, we consider the effect of longitudinal field $h_{z}$. The Hamiltonian in Eq.~\eqref{eq:eq2}  is changed to:
\begin{equation}
	\begin{aligned}
		H_{l}&=-K\sum_{i}A_{i}-K\sum_{i}B_{i}-h_{z}\sum_{i}\tau^{z}_{i}-h_{z}\sum_{i}\sigma^{z}_{i}\,.
		\label{eq:eq4}
	\end{aligned}
\end{equation}
The insertions of $h_{z}$ would actually break the subsystem symmetry as it does not commute with the symmetry generators. Therefore, the SSPT order is immediately broken when $h_{z}>0$~\cite{Phase2009Stein,Fate2012Kalis,Detecting2019Stephen}. Here, we have again measured the energy density $\langle e \rangle$, and the magnetization $\langle m_{\tau^{z}} \rangle$ for the longitudinal field with system size $L=8$, and the results are shown in Fig.~\ref{fig:fig_normal_result_2D_hz}.

From Figs.~\ref{fig:fig_normal_result_2D_hz}(a-b), one sees there is no phase transition at any finite $h_{z}$ and the values of $\langle e \rangle$ and $\langle m_{\tau^{z}} \rangle$ smoothly change as predicted and measured in Ref.~\cite{Fate2012Kalis,Detecting2019Stephen}. In addition, since $h_{z}$ breaks the SSPT order, $\langle O_{d} \rangle$ decay to zero as the membrane size $d$ enlarging for both values of $h_z$ as shown as in Fig.~\ref{fig:fig_normal_result_2D_hz}(c). What's more, setting the membrane size $d=L/2$ with $L$ for the system size, $\langle O_{L/2} \rangle$ also rapidly decays as $h_{z}$ increases. The behavior of $\langle O_{d} \rangle$ means that the SSPT order is indeed immediately destroyed by the insertion of $h_{z}$ field that violates symmetry.

 \subsection{Construction of a class of strange correlators}
\label{sec:StrangCorr}
 Proposed in Refs.~\cite{Wave2014You}, the strange correlator can be roughly understood as a quantity about local operator $\phi(i)$ overlapped by a symmetric trivial direct-product state $|\Omega\rangle$ and a short range entangled state $|\Psi\rangle$ to be diagnosed, which is given by
\begin{equation}
	\begin{aligned}
		C_{\phi}(\Delta r)&=\frac{\langle\Omega|\phi(i+\Delta r)\phi(i)|\Psi\rangle}{\langle\Omega|\Psi\rangle},
		\label{eq:eq6}
	\end{aligned}
\end{equation}
where $\Delta r$ is a two dimensional vector $\Delta r=(\Delta r_{x},\Delta r_{y})$ for the 2D cluster model. The trivial state $|\Omega\rangle$ should hold all the symmetry of $|\Psi\rangle$. In the framework of the projector QMC method \cite{Loop2010Sandvik,Ground2005Sandvik,Computational2010Sandvik},the ground state of the Hamiltonian Eq.~\eqref{eq:eq2} is projected out via an operator string with $n$ bond operators, $|\Psi\rangle=(-H)^{n}|\Psi(0)\rangle$. The strange and normal correlators can respectively be computed as
\begin{equation}
	\begin{aligned}
		C_{\phi}(\Delta r)&=\frac{\langle\Omega|\phi(i+\Delta r)\phi(i)(-H)^{2n}|\Psi(0)\rangle}{\langle\Omega|(-H)^{2n}|\Psi(0)\rangle},
		\label{eq:eq7}
	\end{aligned}
\end{equation}
and
\begin{equation}
	\begin{aligned}
		C^{n}_{\phi}(\Delta r)&=\frac{\langle\Psi(0)|(-H)^{n}\phi(i+\Delta r)\phi(i)(-H)^{n}|\Psi(0)\rangle}{\langle\Psi(0)|(-H)^{2n}|\Psi(0)\rangle},
		\label{eq:eq8}
	\end{aligned}
\end{equation}
where the superscript ``n'' is for the normal correlation.

From Eqs.~\eqref{eq:eq7} and \eqref{eq:eq8}, in the imaginary time setting of the projector QMC simulations, the strange correlator $C_{\phi}(\Delta r)$ reflects the correlation property at the imaginary time boundary between the states $|\Omega\rangle$ and $|\Psi\rangle$~\cite{Wave2014You}. 
Therefore, for $\phi(i)$ defined as a local operator, the strange correlator $C_{\phi}(\Delta r)$ will either saturate to a constant or decay as a power-law in the limit of $\Delta r\rightarrow\infty$ if $|\Psi\rangle$ is a non-trivial SPT state, corresponding to the spatial interfaces between the trivial and non-trivial SPT phases~\cite{Wave2014You}.  Considering the 2D cluster model case, the non-trivial SSPT state $|\Psi\rangle$ is taken to be the ground state of the 2D cluster model, which is projected out with a projection length proportional to the system size (see Appendix~\ref{sec:numerical method} for detailed explanation). Then, the trivial state $|\Omega\rangle$ is preferred to be
\begin{equation}
	\begin{aligned}
		|\Omega\rangle&=\prod_{i} |\tau^{x}_{i,+}\rangle\otimes|\sigma^{x}_{i,+}\rangle\\
		&=\prod_{i} \frac{1}{\sqrt{2}}[|\tau^{z}_{i,+}\rangle+|\tau^{z}_{i,-}\rangle]\otimes|\sigma^{x}_{i,+}\rangle\,,
		\label{eq:eq9}
	\end{aligned}
\end{equation}
where $|\tau^{x}_{i,+(-)}\rangle$ refers to the state that spin $\tau_{i}$ is pointing along (against) the $x$-direction, $|\tau^{z}_{i,+(-)}\rangle$ and $|\sigma^{x}_{i,+(-)}\rangle$ are similarly defined. Note that the state $|\Omega\rangle$ can be viewed as the ground state at the infinite external field $h_{x}$.
In the $\{\tau_{i}^{z}\sigma_{i}^{x}\}$ basis, the state $|\tau^{x}_{i,+}\rangle\otimes|\sigma^{x}_{i,+}\rangle$ can be express as $\frac{1}{\sqrt{2}}[|\tau^{z}_{i,+}\rangle+|\tau^{z}_{i,-}\rangle]\otimes|\sigma^{x}_{i,+}\rangle$. Thus, at the boundary between $|\Omega\rangle$ and $|\Psi\rangle$, flipping $\tau$ spins does not change the weight of given configuration but flipping $\sigma$ spins would lead to zero weight. In another word, within the framework of the projector QMC method, $C_{\phi}(\Delta r)$ is actually about the correlation properties at the boundary of an operator string with a particular boundary condition that all the $\tau$ spins are free to change but the $\sigma$ spins are pinned. However, such a condition causes low efficiency in our projector QMC simulation because all the clusters that touch a $\sigma$ spin are rejected to be flipped and the operator string in our projector QMC simulation is trapped in a local minimum configuration. To overcome it, within the consideration of the subsystem symmetry generators in the 2D cluster model, we also apply the update process that flipping all the $\tau$ spins along a straightforward line parallel to the $x$- or $y$-direction (see Appendix~\ref{sec:Strange correlator} for further details).

From the theoretical side, within the symmetry between $\tau$ and $\sigma$ spin, the strange correlator of the following operators, including the single spin like $\phi(i)=\tau^{z}_{i}$ and $\phi(i)=\sigma^{x}_{i}$, the dimer operator like $D_{i}=\tau_{i}^{z}\tau_{i+\hat{y}}^{z}$, and the multi-spin operator like $\phi(i)=B_{i}$, can be considered. And because a well-established criteria for the optimal choice of $\phi$ operator has not been achieved especially for SSPT case, before the numerical exploration, it is beneficial to obtain some theoretical expectation about the different choices. At the exactly solvable point ($h_{x}=0$), we can straightforwardly obtain these strange correlators   listed in Table~\ref{table:sc_strong}.
In the 2D cluster model,  the strange correlator of the single spin are independent of the transverse field $h_{x}$. For instance, the strange correlator about $\sigma^{x}_{i}$ is always $1$ regardless of $h_{x}$, so such strange correlators are not useful in detecting SSPT order. The numerical calculation and discussion of the strange correlator about the single spin are collected in   Appendices~\ref{sec:Strange correlator} and \ref{sec:exact_calculation}. In summary, in the main text, we will focus on the strange correlators of both $D_{i}$ and $B_{i}$   whose behaviors under the field $h_{x}$ can be used to efficiently detect SSPT order. Specially, for the $D_{i}$ case, we will show that due to its anisotropic behavior (to be demonstrated in Sec.~\ref{sec:results}), the direction of strange correlator also matters in the detection.

\begin{table}
\caption{\textbf{Strange correlators of the 2D cluster model at the exactly solvable point.}
All strange correlators except for $\phi = \tau_{i}^{z}\tau_{i+\hat{y}}^{z},\sigma_{i}^{z}\sigma_{i+\hat{y}}^{z}$ case are constants at the exactly solvable point (i.e. $h_{x}=0$). Here $\hat{x}$ and $\hat{y}$ respectively denote the unit vector along $x$ and $y$ directions shown in Fig.~\ref{fig:fig_model}.
 }
	\begin{tabularx}{8.5cm}{c|c}
		\hline
		
		\hline
		$\phi$& ~~~~~~~~~~~~$C^{}_{\phi}(\Delta r)$ \tabularnewline
		\hline
		
		\hline
		$\tau^{z}_{i}$,$\sigma^{z}_{i}$& ~~~~~~~~~~~~$0$ \tabularnewline
		\hline
		$\sigma^{x}_{i}$,$\tau^{x}_{i}$& ~~~~~~~~~~~~$1$ \tabularnewline
		\hline
		$D_i=\tau_{i}^{z}\tau_{i+\hat{y}}^{z}$, $D'_i=\sigma_{i}^{z}\sigma_{i+\hat{y}}^{z}$&~~~~~~~~~~~~ $1$ for $\Delta r_{y}=0$, $0$ for $\Delta r_{y}\neq0$ \tabularnewline
		\hline
		$B_{i}$,$A_{i}$& ~~~~~~~~~~~~$1$ \tabularnewline
		\hline
		
		\hline
	\end{tabularx}
\label{table:sc_strong}
\end{table}

From the numerical side, with the projector QMC simulation in the $\{\tau^{z}_{i}\sigma^{x}_{i}\}$ basis, the measurement about operators $B_{i}$, $\tau^{z}_{i}$, $\sigma^{x}_{i}$ and $D_i=\tau^{z}_{i} \tau^{z}_{i+\hat{y}}$ is diagonal, which can be directly observed in a given operator string. We present the results    in Sec.~\ref{sec:results} and Appendix \ref{sec:Strange correlator}. However, the measurement about $A_{i}$, $\tau^{x}_{i}$, and $\sigma^{z}_{i}$ is off-diagonal and hard to be measured in the projector QMC simulation directly [see Table~\ref{table:sc_strong}]. But due to the symmetry  between the $\tau$ and $\sigma$ spins, the strange correlator of $A_{i}$, for instance, actually has the same behavior as that of $B_{i}$. Thus, we do not need to simulate the strange correlators of  $A_i$ any more.
 
\subsection{Projector QMC simulation on  strange correlators and strange order parameters}
\label{sec:results}

In the following, we present our numerical results of strange correlator in the 2D cluster model. Firstly, we consider  the strange correlator $C^{}_{\phi}$ where  $\phi$ is $D_i$ in Table~\ref{table:sc_strong}.
\begin{equation}
	\begin{aligned}
		\phi(i) &=D_{i}=\tau^{z}_{i} \tau^{z}_{i+\hat{y}}\\
		C_{D}(\Delta r)&=\frac{\langle\Omega|D_{i+\Delta r} D_{i}(-H)^{2n}|\Psi(0)\rangle}{\langle\Omega|(-H)^{2n}|\Psi(0)\rangle}.
	\end{aligned}
	\label{eq:eq10}
\end{equation}
 
At the exactly solvable point $h_{x}=0$, $C_D(\Delta r)$ would be $1$ for $\Delta r=(\Delta r_{x},0)$, but $0$ for $\Delta r_{y}\neq0$ as demonstrated in Table~\ref{table:sc_strong} and Appendix~\ref{sec:exact_calculation}. From the perspective of symmetries, such an anisotropy is due to the fact that a $D$ operator only transforms non-trivially under certain subsystem symmetries, that makes the behavior of $D_{i+\Delta r} D_{i}$ under symmetry transformations rather complicated. More specifically, for operator $D_i$, we consider two kinds of subsystem symmetry generators $\mathcal{U}_x=\prod_{i\in l_x} \tau^x_i$ and $\mathcal{U}_y=\prod_{i\in l_y} \tau^x_i$, where $l_x$ ($l_y$) is a straight line along $x$- ($y$-) direction. We can see that, for $\mathcal{U}_x$, if $l_x$ contains site $i$ or $i+\hat{y}$, then $\mathcal{U}_x$ anticommutes with $D_i$, as $\mathcal{U}_x$ and $D_i$ share exactly one spin, thus $\mathcal{U}_x D_i \mathcal{U}^{\dagger}_x=-D_i$ (here $\mathcal{U}^{\dagger}$ means the Hermitian conjugate of $\mathcal{U}$, and for $\mathcal{U}_x$ and $\mathcal{U}_y$, we simply have $\mathcal{U}^\dagger=\mathcal{U}=\mathcal{U}^{-1}$); otherwise, $\mathcal{U}_x$ commutes with $D_i$, as $\mathcal{U}_x$ does not share any spin with $D_i$, thus $\mathcal{U}_x D_i \mathcal{U}^{\dagger}_x=D_i$. Meanwhile, $\mathcal{U}_y$ always commutes with $D_i$, as $\mathcal{U}_y$ and $D_i$ share either zero or two spins, thus $\mathcal{U}_y D_i \mathcal{U}^{\dagger}_y=D_i$ always holds.

Then, we consider the behavior of $D_{i+\Delta r} D_i $ under symmetry transformations. Firstly, taking $\Delta r=(\Delta r_{x},0)$, the four $\tau$ spins in $D_{i+\Delta r} D_i $ form the four corners of a membrane. Therefore, $D_{i+\Delta r} D_i$ shares either zero or two spins with any $\mathcal{U}_x$ or $\mathcal{U}_y$ operator, thus they always commute. Since we have $\sigma^{x}_{i}=1$ for all $\sigma$ spins at the trivial state $|\Omega\rangle$, one can see that, for example $\Delta r_{x}>0$, then
\begin{align}
	\langle \Omega|D_{i+\Delta r} D_i  =\langle \Omega| D_{i+\Delta r} \prod_{k\in S} \sigma^x_k D_i =\langle \Omega| \prod_{k\in S} B_k\,,
\end{align} where $S$ is a straight string composed of $\sigma$ spins connecting sites $i$ and $i+\Delta r-\hat{x}$. Therefore, we have $\langle \Omega| D_{i+\Delta r} D_i  |\Psi\rangle =\langle \Omega| \prod_{k\in S} B_k |\Psi\rangle =\langle\Omega|\Psi\rangle$ for a ground state $|\Psi\rangle$ at the exactly solvable point. Consequently, $C_D(\Delta r_x,0)=1$ at $h_{x}=0$.

The case that $\Delta r=(\Delta r_{x},\Delta r_{y})$ and $\Delta r_{y}\neq0$ is different. Since there are at least two $\tau$ spins in the operator $D_{i+\Delta r} D_{i}$ satisfying that for each one of them, its $y$-coordinate is different from the other three $\tau$ spins, one can take one of their locations as site $m$. Then, for symmetry transformation $\mathcal{U}_x=\prod_{i\in l_x} \tau^x_i$ with $m\in l_x$, since $D_{i+\Delta r} D_{i}$ and $\mathcal{U}_x$ only share one spin, we have $\mathcal{U}_x D_{i+\Delta r} D_{i} \mathcal{U}^{\dagger}_x=-D_{i+\Delta r} D_{i}$.
Noticing that both $|\Omega\rangle$ and $|\Psi\rangle$ are invariant under $\mathcal{U}_x$, one can obtain that
\begin{align}
	\!\!\!\!\langle \Omega| D_{i+\Delta r} D_i  |\Psi\rangle\! =\! \langle \Omega| \mathcal{U}_x D_{i+\Delta r} D_i  \mathcal{U}^{\dagger}_x|\Psi\rangle\!=\!-\langle \Omega| D_{i+\Delta r} D_i  |\Psi\rangle\,.
\end{align} As a result, $\langle \Omega| D_{i+\Delta r} D_i  |\Psi\rangle=0$. In conclusion, within the SSPT phase, we expect $C_{D}(\Delta r)$ to decay to nonzero constants only along $x$-direction, which is the direction of $\mathcal{U}_x$ subsystem symmetry generators of 2D cluster model, but no correlation for the others.

\begin{figure}[htp!]
	\centering
	\includegraphics[width=\columnwidth]{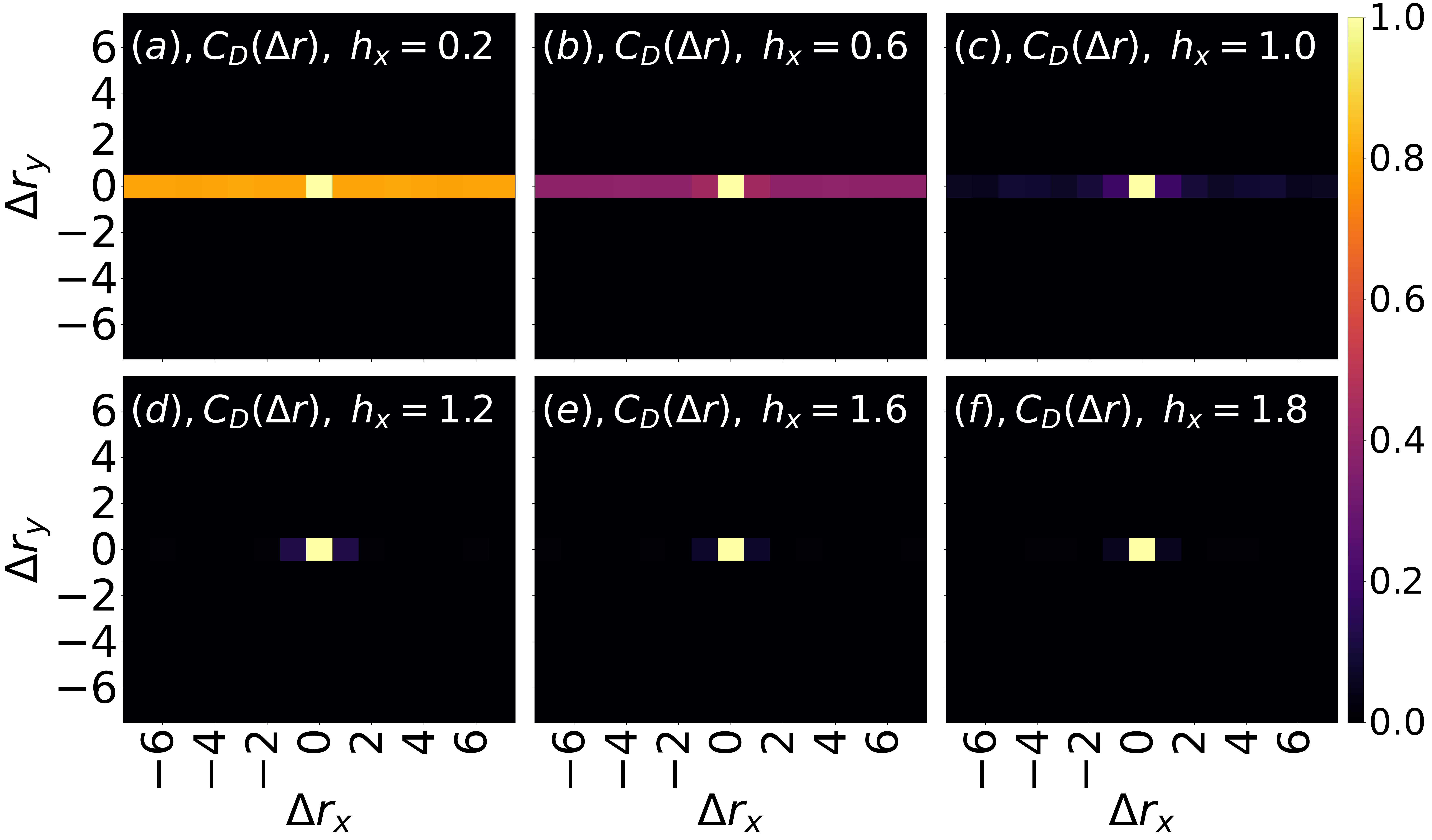}
	\caption{The real-space strange correlator $C_{D}(\Delta r)$ in the 2D cluster (strong SSPT) model. Panel (a-c) are in the strong SSPT phase with $h_{x}$ changing from $0.2$, to $0.6$, and finally to $1.0$, while the others stand for the trivial paramagnetic phase with $h_{x}$ varying from $1.2$ to $1.8$.}
	\label{fig:fig_real_C_D}
\end{figure}

Fig.~\ref{fig:fig_real_C_D} is the real space strange correlator $C_{D}$, where our simulations are carried out with the system size $L=16$. As the field $h_{x}$ increases, the 2D cluster model experiences a phase transition from the SSPT phase into the trivial paramagnetic phase, whose critical point is located at $h_{x,c}=1$ as discussed in Fig.~\ref{fig:fig_normal_result_2D_hx}. Fig.~\ref{fig:fig_real_C_D}(a-c) are $C_{D}(\Delta r)$ measured in the SSPT phase with the transverse field $h_{x}$ changing from $h_{x}=0.2$ to $1.0$ annealing from $h_{x}=0.0$, while (d-f) are that in the trivial paramagnetic phase scanning from $2.0$. In the SSPT phase, along $x$-direction, $C_{D}$ decay to nonzero constants. And along the other directions, like $y$-direction, $C_{D}$ is exactly zero, which means there are no correlation along the other directions.

\begin{figure}[htp!]
\centering
\includegraphics[width=0.95\columnwidth]{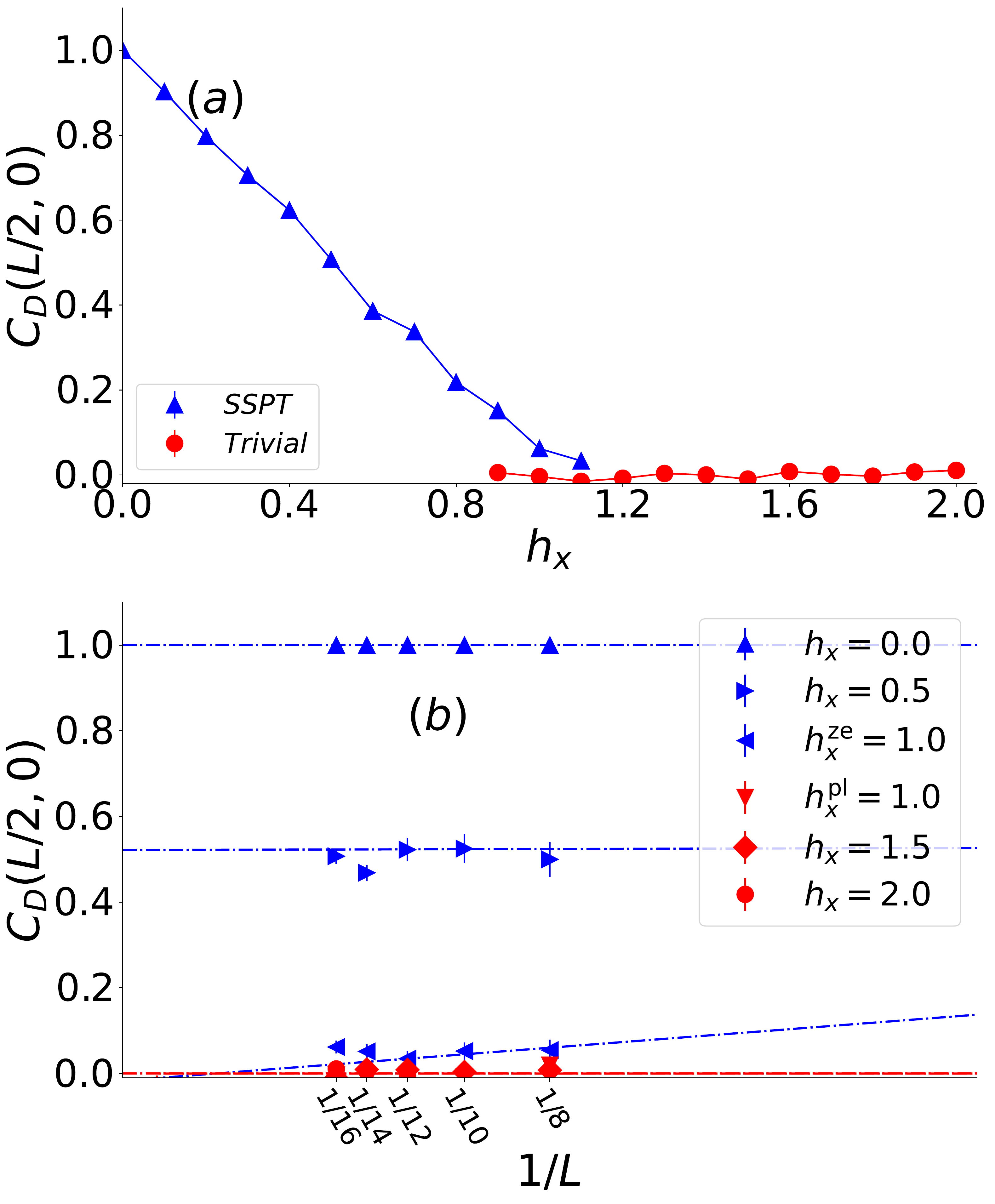}
\caption{The ``strange" order parameter $C_{D}(L/2,0)$. Panel (a) are $C_{D}(L/2,0)$ plotted as a function of $h_{x}$. Panel (b) is the finite-size extrapolation of $C_{D}(L/2,0)$, in which $C_{D}(L/2,0)$ is fitted with $C_{D}(L/2,0)(L=\infty)+a/L$. Here, $h^{\mathrm{ze}}_{x}=1.0$ are annealed from $h_{x}=0.0$ while $h^{\mathrm{pl}}_{x}=1.0$ annealed from $h_{x}=2.0$ in the projector QMC simulations.
}
\label{fig:fig_strangr_op_C_D}
\end{figure}

To further characterize the SSPT order using the anisotropic strange correlator $C_{D}$, a ``strange'' order parameter with the corresponding anisotropy can be introduced as $C_{D}(L/2,0)$, which is $C_{D}$ at the longest distance in the periodic lattice. In 1D case, a similarly defined ``strange" order parameter has been proposed in Ref.~\cite{Ellison2021symmetryprotected}. Fig.~\ref{fig:fig_strangr_op_C_D}(a) shows that $C_{D}(L/2,0)$ decreases as $h_{x}$ increasing. And we also apply the finite-size extrapolation in Fig.~\ref{fig:fig_strangr_op_C_D}(b) of $C_{D}(L/2,0) = C_{D}(L/2,0)(L=\infty)+a/L$ where $L$ is the linear system size to detect the thermodynamic limit behavior of this ``strange" order parameter.
For $h_{x}=0.0, 0.5$, the 2D cluster model stays in the SSPT phase and the extrapolated $C_{D}(L/2,0)(\infty)$ are finite. In contrast, at the transition point and inside the paramagnetic phase, i.e. $h_x \ge h_{x,c}=1.0$, $C_{D}(L/2,0)(\infty)$ clearly extrapolates to zero.

\begin{figure}[htp!]
	\centering
	\includegraphics[width=\columnwidth]{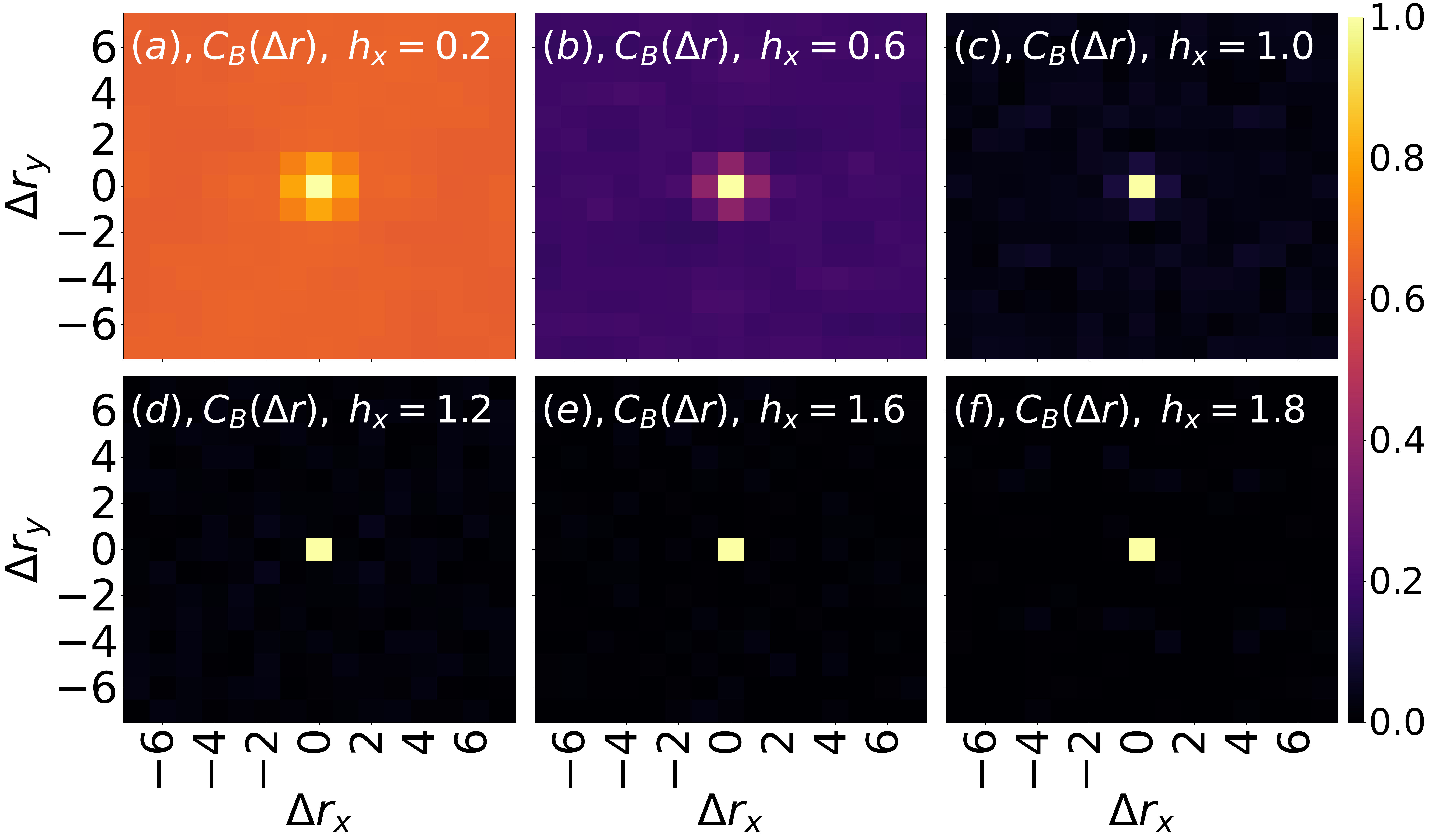}
	\caption{The real-space strange correlator $C_{B}(\Delta r)$ in the 2D cluster (strong SSPT) model. Panel (a-c) are in the strong SSPT phase with $h_{x}$ changing from $0.2$, to $0.6$, and finally to $1.0$, while the others stand for the trivial paramagnetic phase with $h_{x}$ varying from $1.2$ to $1.8$.}
	\label{fig:fig_real_C_B}
\end{figure}

\begin{figure}[htp!]
\centering
\includegraphics[width=0.95\columnwidth]{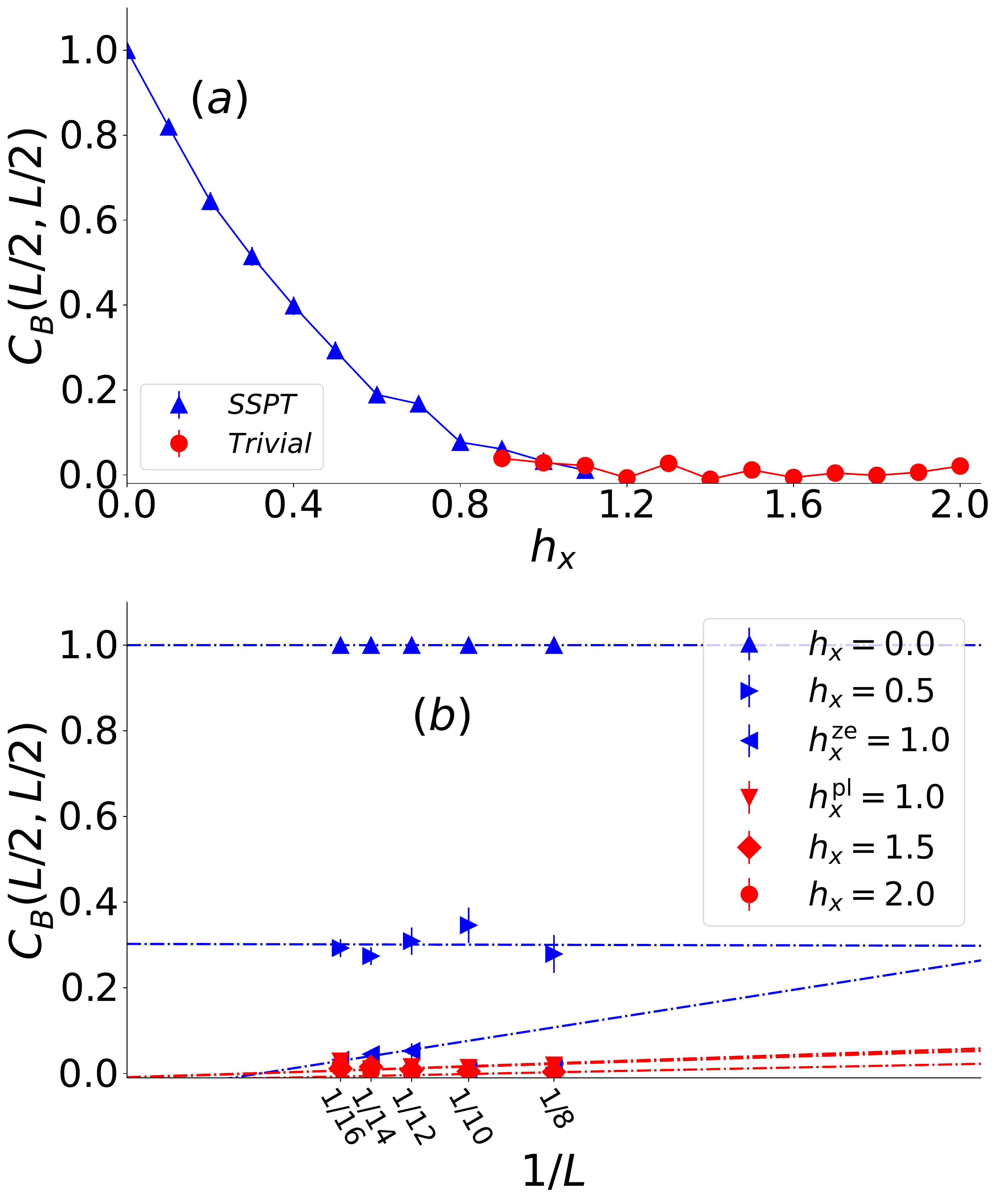}
\caption{The ``strange" order parameter $C_{B}(L/2,L/2)$. Panel (a) are $C_{B}(L/2,L/2)$ plotted as a function of $h_{x}$. Panel (b) is the finite-size extrapolation of $C_{B}(L/2,L/2)$, in which $C_{B}(L/2,L/2)$ is fitted by $C_{B}(L/2,L/2)(L=\infty)+a/L$. Here, $h_{x}=1.0(\mathrm{ze})$ are annealed from $h_{x}=0.0$ while $h_{x}=1.0(\mathrm{pl})$ from $h_{x}=2.0$.}
	\label{fig:fig_strangr_op_C_B}
\end{figure}

Then, we consider the $C_{B}$ with $\phi=B_{i}$ as another  ``strange'' order parameter to describe the SSPT order, where
\begin{equation}
	\begin{aligned}
		\phi(i) &=B_{i}\\
		C_{B}(\Delta r)&=\frac{\langle\Omega|B_{i+\Delta r} B_{i}(-H)^{2n}|\Psi(0)\rangle}{\langle\Omega|(-H)^{2n}|\Psi(0)\rangle}.
	\end{aligned}
	\label{eq:eq11}
\end{equation}
Different from the $C_{D}$ case, $C_{B}$ is isotropic at the exactly solvable point $h_{x}=0.0$, and $B$ transforms trivially under all subsystem symmetries. Our projector QMC data of $C_B$ are shown in Fig.~\ref{fig:fig_real_C_B}. Fig.~\ref{fig:fig_real_C_B}(a-c) tells the real space $C_{B}$ inside the SSPT phase, while (d-f) are that in the paramagnetic phase. One sees $C_{B}$ decay to a constant within the SSPT phase but zero in the paramagnetic phase. Similar with $C_D$, we also use the $C_{B}(L/2,L/2)$ as the ``strange" order parameters and observe how it evolves as $h_{x}$ increasing, which are presented in Fig.~\ref{fig:fig_strangr_op_C_B}.

Fig.~\ref{fig:fig_strangr_op_C_B}(a) describes the dependence of $C_{B}(L/2,L/2)$ with respect to $h_{x}$, which indeed decays to zero as the SSPT order transit to paramagnetic phase at $h_{x,c}$. Similar with the $C_D(L/2,0)$, we also applied the finite-size extrapolation of $C_{B}(L/2,L/2) = C_{B}(L/2,L/2)(L=\infty)+a/L$, which is shown in Fig.~\ref{fig:fig_strangr_op_C_B}(b). As expected, in the thermodynamic limit $C_{B}(\infty)$ is finite inside the SSPT phase ($h_{x}=0,0.5$) while it is zero for the transition point and inside the paramagnetic phase.

\subsection{Projector QMC simulation on the 1D cluster model (Weak version of 2D SSPT)}
\label{sec:1d_case}

Beside the SSPT in the 2D cluster model, there is a weak version of the 2D SSPT, which is adiabatically connected to a decoupled stack of 1D SPT orders without breaking any symmetry. To investigate the strange correlator in this weak SSPT, we also study the 1D cluster model with SPT order protected by global symmetries, whose Hamiltonian can be given as (see Fig.~\ref{fig:fig_model}(c))

\begin{equation}
	\begin{aligned}
		H_{1D}&=-\Gamma\sum_{i}X_{i}\prod_{j}Z_{j}\\
		&=-\Gamma\sum_{i}A^{1D}_i-\Gamma\sum_{i}B^{1D}_i-h_{x}\sum_{i}\tau^{x}_{i}-h_{x}\sum_{i}\sigma^{x}_{i},
		\label{eq:eq5}
	\end{aligned}
\end{equation}
Here, we also define $A^{1D}_i=\tau^{x}_{i}\prod_{j}\sigma^{z}_{j}$ and $B^{1D}_i=\sigma^{x}_{i}\prod_{j} \tau^{z}_{j}$. Different from the $A_{i}$ and $B_{i}$ in the 2D cluster model, $A^{1D}_i$ and $B^{1D}_i$ are three-spin-coupling terms. Also, we take $\Gamma=1$ in the following discussion of the 1D cluster model.
When $h_x=0$, 1D cluster model is exactly solvable because all $A^{1D}_i$ and $B^{1D}_i$ operators commute with each other. Therefore, the ground state of 1D cluster model, $|\Psi\rangle$, satisfies $A^{1D}_i|\Psi\rangle=|\Psi\rangle,\ \forall i$ and $B^{1D}_i|\Psi\rangle=|\Psi\rangle,\ \forall i$, meaning that $|\Psi\rangle$ can be obtained in the same manner as the 2D cluster model (see Appendix~\ref{appendix_gs}). And the quantum critical point in 1D cluster model caused by the external field $h_{x}$ is also located at $h^{1D}_{x,c}=1$~\cite{Three-Spin2004Pachos}.

\begin{figure}[htp!]
	\centering
	\includegraphics[width=\columnwidth]{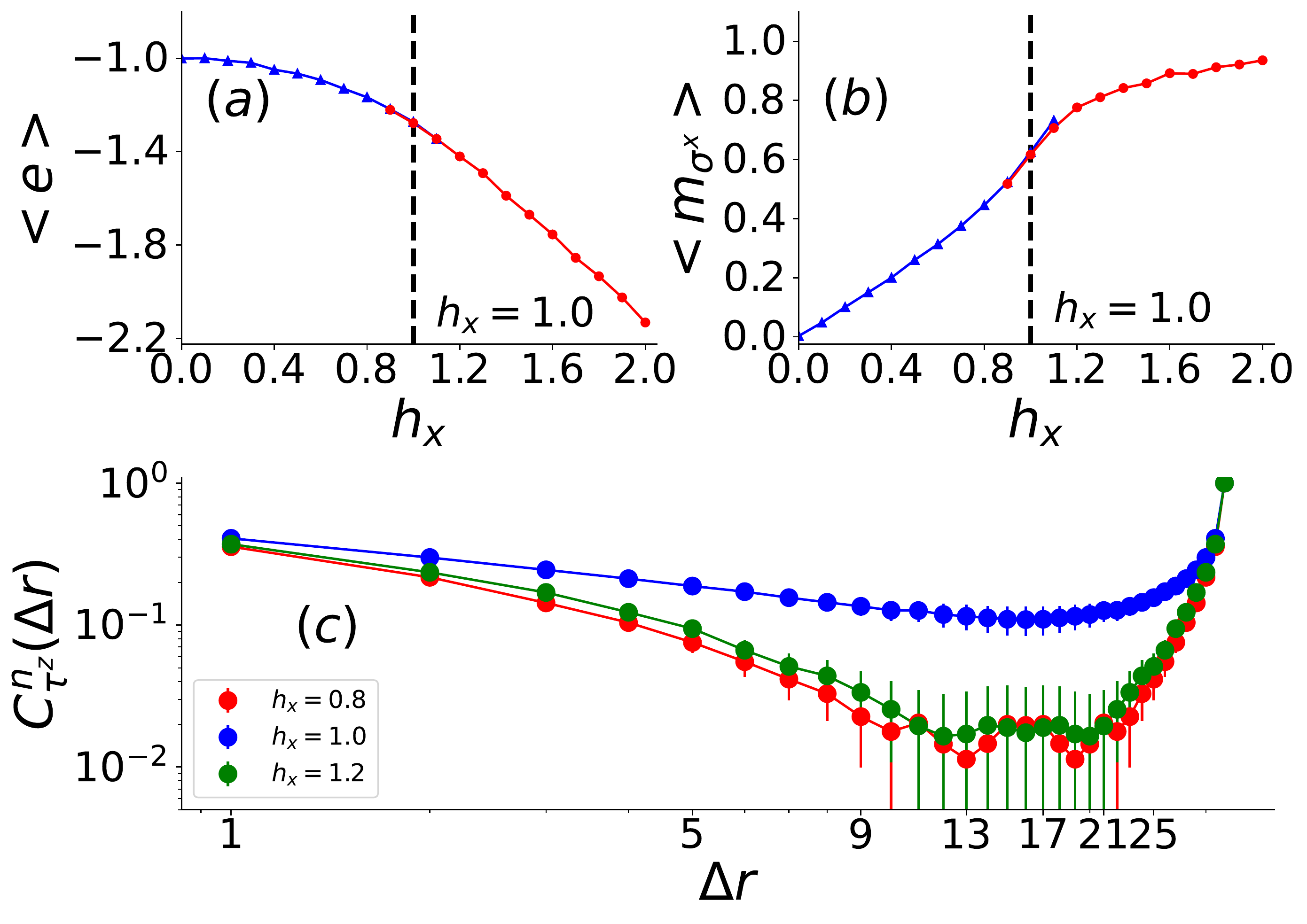}
	\caption{The values of different physical observables under the transverse field $h_{x}$. (a) and (b) are respectively the energy per site and the magnetization of $\sigma$ spins. Among them, the blue lines with triangles are annealing from the exactly solvable point $h_{x}=0$ while the red lines with dots are from the strong field limit $h_{x}=2$. (c) tells the spin-spin correlation functions at $h_{x}=0.8$, $1.0$ , and $1.2$.}
	\label{fig:fig_normal_result_1D_hz}
\end{figure}

With the help of the projector QMC simulation, we have measured the energy density $\langle e \rangle$, the magnetization $\langle m_{\sigma^{x}} \rangle$ and the normal spin-spin correlation $C^{n}_{\tau^{z}}(\Delta r)=\langle\tau^{z}_{i}\tau^{z}_{i+\Delta r}\rangle$ for the 1D cluster model, shown in Fig.~\ref{fig:fig_normal_result_1D_hz}. 
Different from the 2D Cluster case, leading by $h_{x}$, both $\langle e \rangle$ and $\langle m_{\sigma^{x}} \rangle$ experiences a continuous quantum phase transition at $h^{1D}_{x,c}=1$. Moreover, the two point correlation $C^{n}_{\tau^{z}}(\Delta r)$ shows a power-law decay at $h^{1D}_{x,c}=1$, which is consistent with the prediction in Ref.~\cite{Three-Spin2004Pachos}.

So as to capture the SPT nature in the 1D cluster model via strange correlators, we similarly set the operator $\phi$ in Eq.~(\ref{eq:eq7}) to be a single $\tau^z$, $\sigma^x$, and $B^{1D}_i$, respectively~\cite{Ellison2021symmetryprotected}. At the exactly solvable point ($h_{x}=0$), all these strange correlators can all be proved to be $1$, which have been listed in Tab.\ref{table:sc_weak}.
\begin{table}
	\caption{\textbf{Strange correlators of the 1D cluster model at the exactly solvable point.} All strange correlators are constants at the exactly solvable point (i.e., $h_{x}=0$).}
		\begin{tabularx}{8.5cm}{c|c}
		\hline
		
		\hline
		
		$\phi$ &~~~~~~~~~~~~~~~~~~~~~~~~~~~~~~~~~~~~ $C^{1D}_{\phi}(\Delta r)$\tabularnewline
		\hline
		
		\hline
		$\tau^{z}_{i}$,$\sigma^{z}_{i}$& ~~~~~~~~~~~~~~~~~~~~~~~~~~~~~~~~~~~~$1$\tabularnewline
		\hline
		$\sigma^{x}_{i}$,$\tau^{x}_{i}$& ~~~~~~~~~~~~~~~~~~~~~~~~~~~~~~~~~~~~$1$\tabularnewline
		\hline
		$B^{1D}_{i}$,$A^{1D}_{i}$& ~~~~~~~~~~~~~~~~~~~~~~~~~~~~~~~~~~~~$1$\tabularnewline
		\hline
		
		\hline
	\end{tabularx}
	\label{table:sc_weak}
\end{table}

\begin{figure}[htp!]
\centering
\includegraphics[width=0.95\columnwidth]{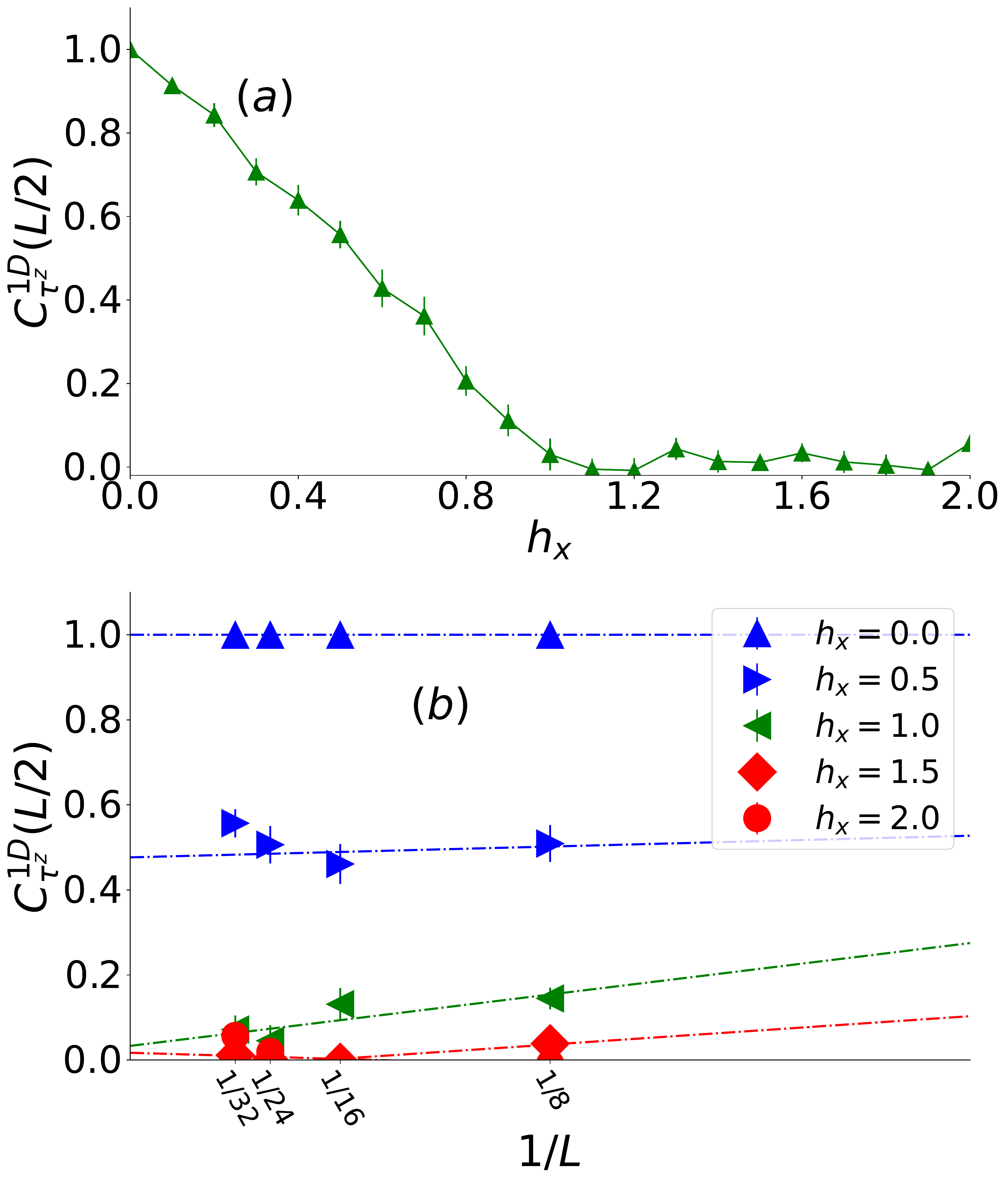}
\caption{The strange order parameter $C^{1D}_{\tau^{z}}(L/2)$. Panel (a) are $C^{1D}_{\tau^{z}}(L/2)$ plotted as a function of $h_{x}$. Panel (b) is the finite-size extrapolation of $C^{1D}_{\tau^{z}}(L/2)=C^{1D}_{\tau^{z}}(L/2)(L=\infty)+c/L$. Inside the 1D SSPT phase, $h_x=0,0.5$, $C^{1D}_{\tau^{z}}(L/2)(L=\infty)$ is finite, whereas at the critical point and inside the paramagnetic phase $C^{1D}_{\tau^{z}}(L/2)(L=\infty)$ is zero.}
	\label{fig:fig_strangr_op_C_t_1D}
\end{figure}

In the main part of this paper, we mainly focus on the most simple form of the strange correlator in the 1D cluster model, which is setting the operator $\phi$ to be a single $\tau^z$ operator
\begin{equation}
\begin{aligned}
\phi(i) &=\tau^{z}_{i}, \\
C_{\tau^{z}}^{1D}(\Delta r)&=\frac{\langle\Omega|\tau^{z}_{i+\Delta r} \tau^{z}_{i}(-H)^{2n}|\Psi(0)\rangle}{\langle\Omega|(-H)^{2n}|\Psi(0)\rangle}.
\end{aligned}
\label{eq:eq12}
\end{equation}
In Fig.~\ref{fig:fig_strangr_op_C_t_1D}(a), we plot the strange correlator $C_{\tau^{z}}^{1D}(L/2)$ as a function $h_x$ for a $L=32$ system and one sees this ``strange" order parameter indeed vanishes at the critical point of $h^{1D}_{x,c}=1$. Also, Fig.~\ref{fig:fig_strangr_op_C_t_1D}(b) is the extrapolation of $C_{\tau^{z}}^{1D}(L) = C_{\tau^{z}}^{1D}(L=\infty)+a/L$. In the SPT (or weak SSPT) phase ($h_{x}=0.0, 0.5$), $C^{1D}_{\tau^z}(\infty)$ is finite. At the quantum critical point $h_{x,c}$ and inside the paramagnetic phase ($h_{x}=1, 1.5,2.0$), $C^{1D}_{\tau^z}(L=\infty)$ vanishes, that is consistent with the phase diagram and our bulk data in Fig.~\ref{fig:fig_normal_result_1D_hz}. In addition, we also measure the strange correlater $C^{1D}_{B}$, which is taking $B^{1D}_{i}$ operator as $\phi$. The numerical result of $C^{1D}_{B}$ is plotted in Fig.\ref{fig:fig_strangr_op_C_B_1D} of Appendix~\ref{sec:Strange correlator}, which also shows the $C^{1D}_{B}$ potential of being a ``strange" order parameter.

\section{Summary and outlook}
\label{sec:discussion}
In this paper, by using the projector QMC simulation within the QA scheme, we have constructed strange correlators of various choices of local operators, and systematically  detected the nontrivial SSPT order and identified the topological phase transition in the 2D cluster model in the presence of transverse magnetic field. In this way, we have successfully transformed the detection of fully localized zero modes on the 1D physical boundary of the SSPT phase to the detection of correlation functions of strange type with the periodic boundary condition, which are very suitable for the large-scale QMC simulation.  More concretely, for the 2D cluster model considered in this paper, the strange correlator $C_{D}(\Delta r)$ at large $\Delta r$ serves as a ``strange" order parameter to sensitively detect the transition between the SSPT phase and the  trivial  paramagnetic  phase. Moreover, $C_{D}$ shows an interesting spatial anisotropy, which can be intrinsically traced back to the nature of spatial anisotropy of subsystem symmetries that protect SSPT order in the 2D cluster model. Meanwhile, $C_B(\Delta r)$ also serves as a sensitive ``strange'' order parameter of the SSPT phase, and together with $C_D$, it is finite inside the SSPT phase and zero in the trivial paramagnetic phase within the errorbar. Our constructions of $C_D$ and $C_B$, exhibit the versatile and easy-to-implementment nature of the strange correlator in studying the SSPT systems. 

While our QMC results demonstrate that the strange correlator diagnosis is powerful in the detection of SSPT orders,  given an SSPT phase, which is protected by subsystem symmetry with infinite number of independent generators in the thermodynamic limit, a general principle to design a strange correlator still needs further exploration and clarification. More specifically, an optimal choice of the local operator in a strange correlator for such a case does not have a well-established criteria yet. Therefore, before numerical exploration, we firstly give a brief theoretical discussion to see which operators can be expected to show non-trivial numerical results (see Sec.~\ref{sec:StrangCorr}). And in our numerical setting, for instance, $D$ operator transforms nontrivially under certain subsystem symmetries, and we can notice that there is a correspondence between $D$ and edge modes. That is, we have $\langle\Omega|D_{i}=\langle\Omega|D_{i} \sigma^x_i=\langle\Omega|\tau^z_i \sigma^x_i \tau^z_{i+\hat{y}}$, where $\tau^z_i \sigma^x_i \tau^z_{i+\hat{y}}$ has the same form with the $\pi^x$ operator of an effective spin on a boundary extended along $y$-direction with a $\sigma$ spin sitting on the center (see Sec.~\ref{sec:Model}). At the same time, $B$ operator transforms trivially under all subsystem symmetries and it does not show a similarly direct correspondence with edge modes, but it can be recognized as a membrane order parameter with size $d=1$ (see Sec.~\ref{sec:phase_diag}). Despite the different symmetry properties, as demonstrated by the numerical results, both $C_D$ and $C_B$ show behavior of strange order parameters of the SSPT phase in the 2D cluster model within the present numerical precision. Recently, some discussion on an optimal choice of local operators in strange correlators of SPT phases protected by global symmetries has been presented in Ref.~\cite{Strange2022Lepori}, while such a type of  discussion on SSPT phases is still lacking. We expect our numerical results will be beneficial to further theoretical studies. 
In addition,   it was assumed that the (quasi-)long-range behavior of strange correlators is related to the spatial interface between SPT phases and trivial phases by applying the theoretical argument of  Lorentz transformations~\cite{Wave2014You}; however, subsystem symmetries are  incompatible with Lorentz invariance~\cite{Generalized2022McGreevy}, and yet our results clearly demonstrate the strange correlators successfully detect the SSPT phase and its transition to trivial phase.   Moreover, it is interesting to build a more direct bridge between more traditional physical observables (e.g., bulk and boundary excitations) of SSPT phases and the finite value of strange correlators at long distances (i.e., the strange order parameter). 
Overall, systematical explorations on the effectiveness of strange correlators as well as the generic theoretical understanding for the construction of strange correlators for topological phases including both SPT and SSPT are clearly posted to the community. Along with the previous   studies   in the topic of strange correlators, we hope all these results  will be helpful in the future in constructing a general theoretical framework of strange correlators.

\acknowledgements
We thank Yi-Zhuang You for helpful discussions. CKZ, ZY and ZYM acknowledge support from the Research Grants Council of Hong Kong SAR of China (Grant Nos.~17303019, 17301420, 17301721, AoE/P-701/20 and 17309822), the K. C. Wong Education Foundation (Grant No.~GJTD-2020-01), and the Seed Funding ``Quantum-Inspired explainable-AI'' at the HKU-TCL Joint Research Centre for Artificial Intelligence. We thank HPC2021 system under the Information Technology Services and the Blackbody HPC system at the Department of Physics, the University of Hong Kong for their technical support and generous allocation of CPU time. MYL and PY were supported by NSFC Grant (No.~12074438). MYL and PY are supported in part by the Guangdong Provincial Key Laboratory of Magnetoelectric Physics and Devices (LaMPad).
\newpage
\appendix
\bibliography{ref}

\section{Numerical method}
\label{sec:numerical method}
Here, we perform our simulation for the 2D cluster model using the projector Quantum Monte Carlo (QMC) method\cite{Ground2005Sandvik,Loop2010Sandvik,Computational2010Sandvik}. Such a approach is based on the expression that
\begin{equation}
	\begin{aligned}
		(-H)^n|\Psi(0)\rangle&=c_{0}(-E_{0})^{n}\left[|0\rangle+\sum_{m=1}^{\Lambda-1}\frac{c_{m}}{c_0}\left(\frac{E_{m}}{E_{0}}\right)^n|n\rangle\right],
	\end{aligned}
	\label{eq:apeq1}
\end{equation}
where state $|n\rangle$ refers to the energy eigenstates in a Hilbert space of $\Lambda$ states. Noted the $(-H)^n|\Psi(0)\rangle\propto|0\rangle$ if $E_{0}$ is the largest eigenvalue and its expansion coefficient $c_{0}\neq0$. Therefore, in the projector QMC method, the ground state $|0\rangle$ can be projected out from an arbitrary trial state $|\Psi(0)\rangle$ by sampling the terms of $(-H)^n|\Psi(0)\rangle$ with a final extrapolation $n\rightarrow\infty$, which construct a sampling configuration space\cite{Ground2005Sandvik,Existence1990Liang}.

Then, under the $\{\tau^{z}_{i}\sigma^{x}_{i}\}$ basis, we first rewrite the Hamiltonian Eq.\ref{eq:eq2} as
\begin{equation}
	\begin{aligned}
		H&=-\sum_{i}H_{A_{i}}-\sum_{i}H_{B_{i}}-\sum_{i}H_{h_{\tau,i}}-\sum_{i}H_{h_{\sigma,i}}+H_{C}.
	\end{aligned}
	\label{eq:apeq2}
\end{equation}
with $H_{A_{i}}=K(A_{i}+I_{5})$, $H_{B_{i}}=K(B_{i}+3I_{5})$, $H_{h_{\tau,i}}=h_{x}(\tau^{x}_{i}+I_{2})$, and $H_{h_{\sigma,i}}=h_{x}(\sigma^{x}_{i}+3I_{2})$. And, $H_{C}=4KI_{5}+4h_{x}I_{2}$, where $I_{n}$ are identity matrix of order $n$. Following the framework of projector QMC method, all of the none-zero elements in the Hamiltonian can be read as $\langle H_{A_{i}} \rangle=K$, $\langle H_{B_{i}} \rangle=2K$ or $4K$, $\langle H_{h_{\tau,i}} \rangle=h_{x}$, and $\langle H_{h_{\sigma,i}} \rangle=2h_{x}$ or $4h_{x}$. In the projector QMC method, the concept of the operator string is introduced by writing from a trial state that
\begin{equation}
	\begin{aligned}
		(-H)^n|\Psi(0)\rangle&=\sum_{\alpha}\prod_{n}^{i=1}H_{i}|\Psi(0)\rangle=\sum_{\alpha}W_{\alpha}|\Psi(\alpha)\rangle.
	\end{aligned}
	\label{eq:apeq3}
\end{equation}
with $H_{i}$ standing for different term in Eq.\ref{eq:apeq2} and $\alpha$ is the formal label for the different strings. $|\Psi(\alpha)\rangle$ denotes the state obtained when the operators acted on $|\Psi(0)\rangle$ and $W_{\alpha}=\prod_{i}\langle H_{i} \rangle$. By sampling a high power of $H$ and its action on trial state $|\Psi(0)\rangle$, the ground state can be projected out.

In our projector QMC simulation, there are two kinds of update, which are the diagonal and the off-diagonal update. They are presented as below.

\subsection{Diagonal update}
\label{sec:appA1}
The diagonal update is about exchanging the operators. Firstly, starting from a initial trial state, a operator string is constructed by randomly selecting $n=32L^3$ operators, where $L$ is the system size. And, in the diagonal update process, we scan the operator string and find the diagonal operators. For each diagonal operators, it would be replaced by a new diagonal operator selected according to the following process. The type of diagonal operator is firstly determined according to the probability
\begin{equation}
	\begin{cases}
		P_{A}=&\frac{KN_{A}}{KN_{A}+4KN_{B}+5h_{x}N_{s}}\\
		P_{B}=&\frac{4KN_{B}}{KN_{A}+4KN_{B}+5h_{x}N_{s}} \\
		P_{h_{\tau}}=&\frac{h_{x}N_{s}}{KN_{A}+4KN_{B}+5h_{x}N_{s}} \\
		P_{h_{\sigma}}=&\frac{4h_{x}N_{s}}{KN_{A}+4KN_{B}+5h_{x}N_{s}}\\
	\end{cases},
	\label{eq:apeq4}
\end{equation}
where $N_{A}$ is the total number of $A_{i}$ in Eq.\ref{eq:apeq2} and $N_{B}$ is that of $B_{i}$. And $N_{s}$ refers to the number of the spin $\tau$ in the 2D cluster model, which is equal to the number of the spin $\sigma$.

If the type $A_{i}$ is picked up, we randomly select a position $s$ from $N_{A}$ and insert a diagonal operator $A_{i}$ with probability
 \begin{equation}
 		P^{insert}_{A}=\frac{\langle H_{A_{s}} \rangle}{K}=1\\
 	\label{eq:apeq5}
 \end{equation}
where $\langle H_{A_{s}} \rangle$ is the expectation value of operator $A_{i}$ at position $s$. Since the only non-zero element of $H_{A_{s}}$ is $K$, the insertion must be accepted.

If the type $B_{i}$ is picked up, a position $s$ is randomly picked up from $N_{B}$ and inserted with probability
\begin{equation}
	P^{insert}_{B}=\frac{\langle H_{B_{s}} \rangle}{4K}\\
	\label{eq:apeq6}
\end{equation}
where $\langle H_{B_{s}} \rangle$ is the expectation value of operator $B_{i}$ at position $s$. If the insertion is rejected, we go back to the selecting operator type process and repeat the inserting process.

If the type $H_{h_{\tau,i}}$ is chosen, spin $\tau_{s}$ is randomly selected from $N_{s}$ and insert a diagonal operator $H_{h_{\tau}}$ with probability
\begin{equation}
	P^{insert}_{h_{\tau}}=\frac{\langle H_{h_{\tau,s}} \rangle}{h}=1\\
	\label{eq:apeq7}
\end{equation}
where $\langle H_{h_{\tau,s}} \rangle$ is the expectation value of operator $H_{h_{\tau,s}}$ at spin $\tau_{s}$. Since the only non-zero element of $H_{h_{\tau_{s}}}$ is $h_{x}$, the insertion must be accepted.

If the type $H_{h_{\sigma,i}}$ is picked up, spin $\sigma_{s}$ is randomly selected from $N_{s}$ and inserted with probability
\begin{equation}
	P^{insert}_{h_{\sigma}}=\frac{\langle H_{h_{\sigma_{s}}} \rangle}{4h}\\
	\label{eq:apeq6}
\end{equation}
where $\langle H_{h_{\sigma,s}} \rangle$ is the expectation value of operator $H_{h_{\sigma,s}}$ at position $\sigma_{s}$. If the insertion is rejected, we go back to the selecting operator type process and repeat the inserting process.

\subsection{Off-diagonal update}
\label{sec:appA2}
For the off-diagonal update process, both the local update and the modified cluster update are applied in our simulation\cite{Stochastic2003Sandvik,Sweeping2019Yan,Global2022Yan}. First of all, there are two kinds of operators in the operator string, which are the pure diagonal operator ($H_{B_{i}}$ and $H_{h_{\sigma,i}}$) and the quantum operator ($H_{A_{i}}$ and $H_{h_{\tau,i}}$). Caused by the constant term in Eq.\ref{eq:apeq2}, the diagonal element in $H_{A_{i}}$ and $H_{h_{\tau,i}}$ are no-zero. Consequently, in the operator string, the quantum operator can be both diagonal and off-diagonal. To achieve a high efficiency, we applied both the local update and the cluster update in the projector QMC simulation.

\begin{figure}[htp!]
	\centering
	\includegraphics[width=\columnwidth]{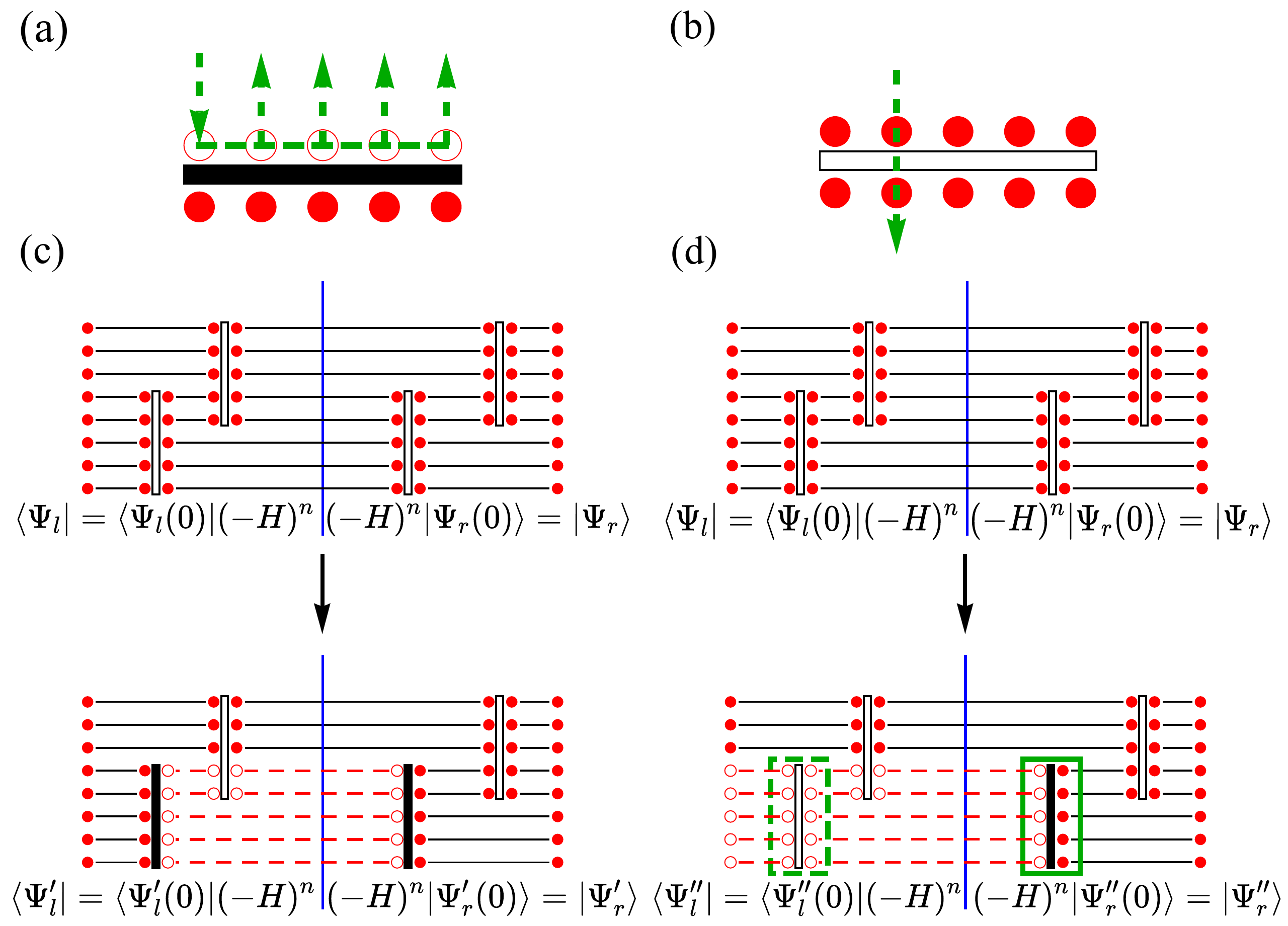}
	\caption{Schematic diagrams of the off-diagonal update. The red dashed line is the region suggested to flip the spin. Under the Metropolis process, its acceptance probability is $P_{fl}=min\left(1,\frac{W_{new}}{W_{old}}\right)$. And (a) and (b) describe two different way when a cluster line meets an operator. (c) tells the local update and (d) refers to the cluster update process in which the operator in the solid rectangle is deal in the (a) way but that in the dashed rectangle is in the (b) way.}
	\label{fig:fig_schematic_update}
\end{figure}

When it comes to the local update process, a leg of a quantum operator ($H_{A_{i}}$ and $H_{h_{\tau,i}}$) is selected randomly in a given operator string. Then, from this leg, we create all the update-lines of this vertex and evolve them along the operator string until it meets another operator acting on the same position (see Fig.~\ref{fig:fig_schematic_update}(c)). When the update line meets the boundary of the operator string ($\langle\Psi_{l}(0)|$ and $|\Psi_{r}(0)\rangle$), the update line would be ended. The spins included in the update region are proposed to be flipped (the red region in the Fig.~\ref{fig:fig_schematic_update}(c)).

For the cluster update, also starting from a randomly selected vertex leg of a quantum operator, the cluster is constructed under the rules listed as follows. (1) When the cluster building line meets a pure diagonal operator ($H_{B_{i}}$ and $H_{h_{\sigma,i}}$), it would go through the operator directly which is presented as Fig.~\ref{fig:fig_schematic_update}(b). (2) When the cluster building line meets a quantum operator ($H_{A_{i}}$ and $H_{h_{\tau,i}}$), it would evolve in two different ways. This line can cross the operator directly as Fig.~\ref{fig:fig_schematic_update}(b). Or it will be ended at its vertex leg but generating update lines from all the other vertex legs (see Fig.~\ref{fig:fig_schematic_update}(a)).  (3) The update line would also be terminated when it meets the boundary ($\langle\Psi_{l}(0)|$ and $|\Psi_{r}(0)\rangle$). In each cluster update process, we pick $1\%$ of these quantum operators randomly and treat them in the Fig.~\ref{fig:fig_schematic_update}(a) way in the cluster constructing process while the others in the Fig.~\ref{fig:fig_schematic_update}(b) ways. Noted that this cluster building process would turn back into the typical cluster update with treating each quantum operators in the Fig.~\ref{fig:fig_schematic_update}(b) way. Finally, the spins including in the cluster (the red region in the Fig.~\ref{fig:fig_schematic_update}(d)), are suggested to be flipped.

In both of these updates, the spins included in the red region would be flipped with the acceptances probability given by
\begin{equation}
	\begin{aligned}
		P_{fl}&=min\left(1,\frac{W_{new}}{W_{old}}\right),\\
		\frac{W_{new}}{W_{old}}&=2^{n^{+,new}-n^{+,old}}.
	\end{aligned}
	\label{eq:eq12}
\end{equation}

Here, $n^{+,new (old)}$ are the number of diagonal operators with $\langle H_{B_{i}(h_{\tau,i})}\rangle=4K(h)$ in the new (old) configuration. It means that the weight ratio depends on the number of the overlap-value-changing diagonal operator.

Moreover, due to the first-order phase transition leaded by $h_{x}$, the QA process is required in our numerical simulation to make a faster convergence to the ground state, in which the quantum parameter, $h_{x}$, would be slowly changed and the operator string from the last parameter result would be applied as a new initial string for projector QMC simulation at the next parameter. Our simulation in the SSPT phase of the 2D cluster model scans from the exactly solvable point with an annealing step $\Delta h_{x}=0.01$ and over $10^4$ Monte Carlo steps at each annealing step\cite{Quantum1998Kadowaki,Theory2002Santoro,Targeting2022Yan}. And the measurements in the paramagnetic phase are from large field limit with the same annealing step.

\subsection{Measurement}
\label{sec:appA3}

In the projector QMC method, to calculate the expectation value of operator $P$ in the ground state $\langle P \rangle$, one can rewrite it in the terms of two projector states,
\begin{equation}
	\begin{aligned}
		\langle P \rangle&=\frac{\langle\Psi_{l}(0)|(-H)^n P (-H)^n|\Psi_{r}(0)\rangle}{\langle\Psi_{l}(0)|(-H)^{2n}|\Psi_{r}(0)\rangle}\\
		&=\frac{\sum_{\alpha_l}\sum_{\alpha_r}W_{\alpha_l}W_{\alpha_r}\langle\Psi_{l}(\alpha)|P|\Psi_{r}(\alpha)\rangle}{\sum_{\alpha_l}\sum_{\alpha_r}W_{\alpha_l}W_{\alpha_r}\langle\Psi_{l}(\alpha)|\Psi_{r}(\alpha)\rangle}.
	\end{aligned}
	\label{eq:eq13}
\end{equation}
Here, the weight function used in important sampling is $W_{\alpha_l}W_{\alpha_r}\langle\Psi_{\alpha_l}|\Psi_{\alpha_r}\rangle$ and the operator estimator is $\langle\Psi_{\alpha_l}|P|\Psi_{\alpha_r}\rangle/\langle\Psi_{\alpha_l}|\Psi_{\alpha_r}\rangle$. Within the picture of the operator string, this measurement is applied at the middle of the operator string.

Moreover, for the ground state energy, a reference state $|R\rangle$ that the equal-amplitude superposition of all spin configuration in the $\{\tau^{z}_{i}\sigma^{x}_{i}\}$ basis is selected, and the ground state energy takes
\begin{equation}
	\begin{aligned}
		E_{0}&=\frac{\langle R|H|0\rangle}{\langle R|0\rangle}\\
		&=\frac{\sum_{\alpha}W_\alpha\langle R|H|\Psi(\alpha)\rangle}{\sum_{\alpha}W_\alpha\langle R|\Psi(\alpha)\rangle}.
	\end{aligned}
	\label{eq:apeq14}
\end{equation}
Here the weight function sampled is $W_{\alpha}\langle R|\Psi_{\alpha}\rangle$ and the operator estimator is $\langle R|H|\Psi_{\alpha}\rangle/\langle R|\Psi_{\alpha_r}\rangle$\cite{Ground2005Sandvik}. Since all the overlaps $\langle R|\Psi\rangle$ keeping the same value, they can be canceled. With the operator string is sampled with probability proportional to $W_{\alpha}$, the energy can be read as
\begin{equation}
\begin{aligned}
	E_{0}&=KN_{A}+K\sum_{i}n^{B,\pm}_{i}+h_{x}N_{s}+h_{x}\sum_{i}n^{\sigma,\pm}_{i},
\end{aligned}
\label{eq:apeq15}
\end{equation}
where $n^{B(\sigma),\pm}_{i}=\pm1$ is the expectation values of operator $\langle R|B_{i}|\Psi(\alpha)\rangle$ and $n^{\sigma,\pm}_{i}=\pm1$ for that of $\langle R|\sigma^x_i|\Psi(\alpha)\rangle$ on the $\sigma$ spin.

\section{Ground state of the 2D cluster model}
\label{appendix_gs}

\begin{figure}[htp!]
	\centering
	\includegraphics[width=0.6\columnwidth]{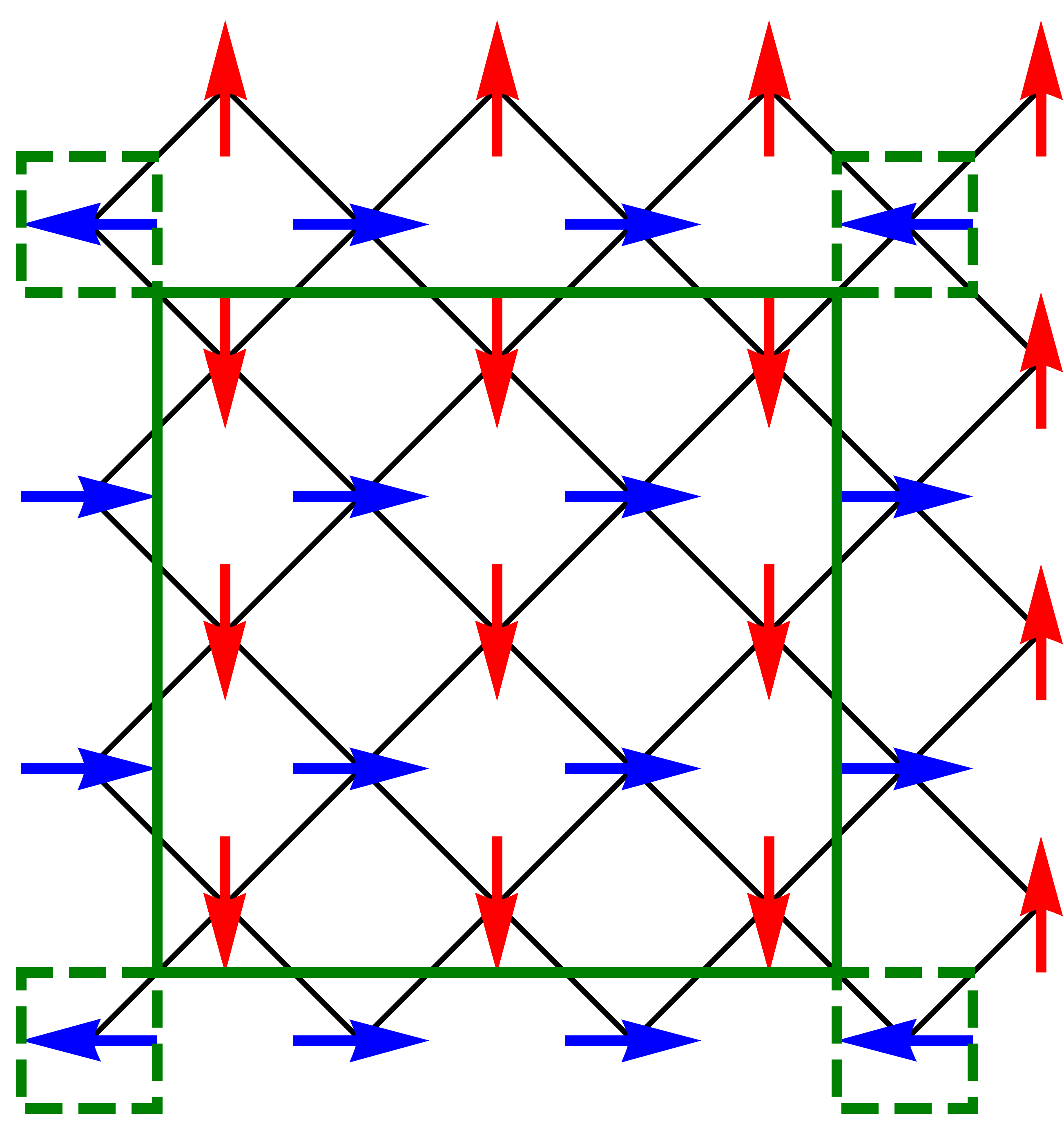}
	\caption{An illustration of the ground state of 2D cluster model. The blue arrows denote the $\tau$ spins, with right arrows for $\tau^x=1$ and left arrows for $\tau^x=-1$. The red arrows denote $\sigma$ spins, with up arrows for $\sigma^z=1$ and down arrows for $\sigma^z=-1$. As we can see, the solid green lines denote the domain walls between $\sigma^z$'s with opposite values, and the dashed green squares denote corners of the domain walls decorated with $\tau^x=-1$.}
	\label{fig:fig_gs}
\end{figure}

In this section we review the unique ground state of 2D cluster model with periodic boundary condition in the exactly solvable point (i.e. $h_x=0$). Without the transverse fields $h_{x}$, the 2D cluster model is exactly solvable. To understand its ground state, it is worth noticing that when $h_x=0$, every term in Eq.~\eqref{eq:eq2} commutes with each other. Consequently, the ground state of the 2D cluster model $|\Psi\rangle$ is the eigenstate of all $A$ and $B$ terms with eigenvalue $1$ (i.e., $A_i|\Psi\rangle=|\tau^{x}_{i}\prod_{j}\sigma^{z}|\Psi\rangle=|\Psi\rangle,\ \forall i$ and $B_i|\Psi\rangle=\sigma^{x}_{i}\prod_{j} \tau^{z}_{i}|\Psi\rangle=|\Psi\rangle,\ \forall i$). Therefore, with periodic boundary condition, we can explicitly construct the unique ground state by the following steps:
\begin{itemize}
	\item First, we take a reference state $|R\rangle$ which is the eigenstate of all $\sigma^z$ and $\tau^x$ operators with eigenvalue $1$. It is obvious that  $A_i|R\rangle = |R\rangle,\ \forall i$. In this section, for convenience, an eigenstate of all $\sigma^z$ and $\tau^x$ operators is dubbed as a configuration. Obviously, such configurations form a complete and orthogonal basis of the Hilbert space of the system.
	\item Then, we can find that, the equal-weight superposition of all configurations that can be obtained by applying $B$ operators on $|R\rangle$ is exactly the ground state $|\Psi\rangle$ (as a $B$ operator always flip the eigenvalues of a $\sigma^z$ and four $\tau^x$ operators, all states that can be obtained by applying $B$ operators on $|R\rangle$ are configurations). To see this, we need to notice that because all $A$ and $B$ operators commute with each other, all configurations that can be obtained by applying $B$ operators on $|R\rangle$ are still eigenstates of all $A$ operators with eigenvalue $1$. And according to the construction of $|\Psi\rangle$, where two configurations that are related by a $B$ operator are always equal-weight superpositioned, $|\Psi\rangle$ is also the eigenstate of all $B$ operators with eigenvalue $1$.
\end{itemize}
By observation, as a $B$ operator can be recognized as flipping a single $\sigma^z$ at the center of a small membrane and the four $\tau^x$ at the four corners of the small membrane, an arbitrary configuration that can be obtained by applying $B$ operators on $|R\rangle$ can be regarded as an Ising configuration of $\{\sigma^z\}$ with $\tau^x=-1$ decorated at the corners of the domain walls between $\sigma^z$'s with opposite values, and $\tau^x=1$ for all other $\tau$ spins. To see this, we only need to notice that for an Ising configuration of $\{\sigma^z\}$, we can regard all down spins (i.e. $\sigma^z=-1$) as being applied by membranes composed of $B$ operators, and the corners of domain walls between $\sigma^z$'s with opposite values are exactly the corners of such membranes, thus they have to contain $\tau^x=-1$ due to the action of $B$ operators. As a result, $|\Psi\rangle$ can described as a superposition of all Ising configurations of $\{\sigma^{z}\}$ with a) $\tau^{x}=-1$ decorated on all corners of domain walls between $\sigma^z$'s with opposite values and b) $\tau^{x}=1$ for all other $\tau$ spins~\cite{Subsystem2018You}. A pictorial demonstration of such a configuration is given in Fig.~\ref{fig:fig_gs}. At last, here it should be noticed that in numerical simulation we use $\{\tau_{i}^{z}\sigma_{i}^{x}\}$ basis. The discussion about the ground state here can also be applied in that basis.

\section{Strange correlator measurment via projector QMC simulation}
\label{sec:Strange correlator}

For the strange correlator $C_{\phi}(\Delta r)$ with chosen trivial state $|\Omega\rangle$, it can be given as
\begin{equation}
	\begin{aligned}
		C_{\phi}(\Delta r)&=\frac{\langle \Omega|C_{\phi}(\Delta r)|0\rangle}{\langle \Omega|0\rangle}\\
		&=\frac{\sum_{\alpha}W_\alpha\langle \Omega|C_{\phi}(\Delta r)|\Psi(\alpha)\rangle}{\sum_{\alpha}W_\alpha\langle \Omega|\Psi(\alpha)\rangle},
	\end{aligned}
	\label{eq:apeq16}
\end{equation}
with the weight function $W_\alpha\langle \Omega|\Psi(\alpha)\rangle$ and estimator $\langle \Omega|C_{\phi}(\Delta r)|\Psi(\alpha)\rangle/\langle \Omega|\Psi(\alpha)\rangle$.

It is worth to note that the measurement here is applied at the boundary between $|\Omega\rangle$ and $|\Psi(\alpha)\rangle$. Choosing the trivial state (Eq.\ref{eq:eq9}) in the strange correlator leads to the particular boundary condition between $|\Omega\rangle$ and $|\Psi(\alpha)\rangle$ in the projector QMC simulation. For instance, taking $|\Omega\rangle=\prod_{i} \frac{1}{\sqrt{2}}[|\tau^{z}_{i,+}\rangle+|\tau^{z}_{i,-}\rangle]\otimes|\sigma^{x}_{i,+}\rangle]$, in which state $|\tau^{z}_{i,\pm}\rangle=\pm1$ share the same amplitude but only $|\sigma^{x}_{i,+}\rangle$ has a non-zero amplitude. Thus, any cluster that flipping spin $\tau^{z}$ at the boundary between $|\Omega\rangle$ and $|\Psi(\alpha)\rangle$ would not change the weight function $W_\alpha\langle \Omega|\Psi(\alpha)\rangle$ while that flipping $\sigma^{x}$ at the boundary causes $W_\alpha\langle \Omega|\Psi(\alpha)\rangle=0$ and is always rejected. Therefore, within the picture of the projector QMC simulation, spins $\sigma$ at the boundary between $|\Omega\rangle$ and $|\Psi(\alpha)\rangle$ is pinned at the state $|\sigma^{x}_{i,+}\rangle$, while spins $\tau$ at the boundary are free to be flipped. And finally, the $ C_{\phi}(\Delta r)$ measurement can be simply applied at the boundary between $|\Omega\rangle$ and $|\Psi(\alpha)\rangle$.

\begin{figure}[htp!]
	\centering
	\includegraphics[width=0.95\columnwidth]{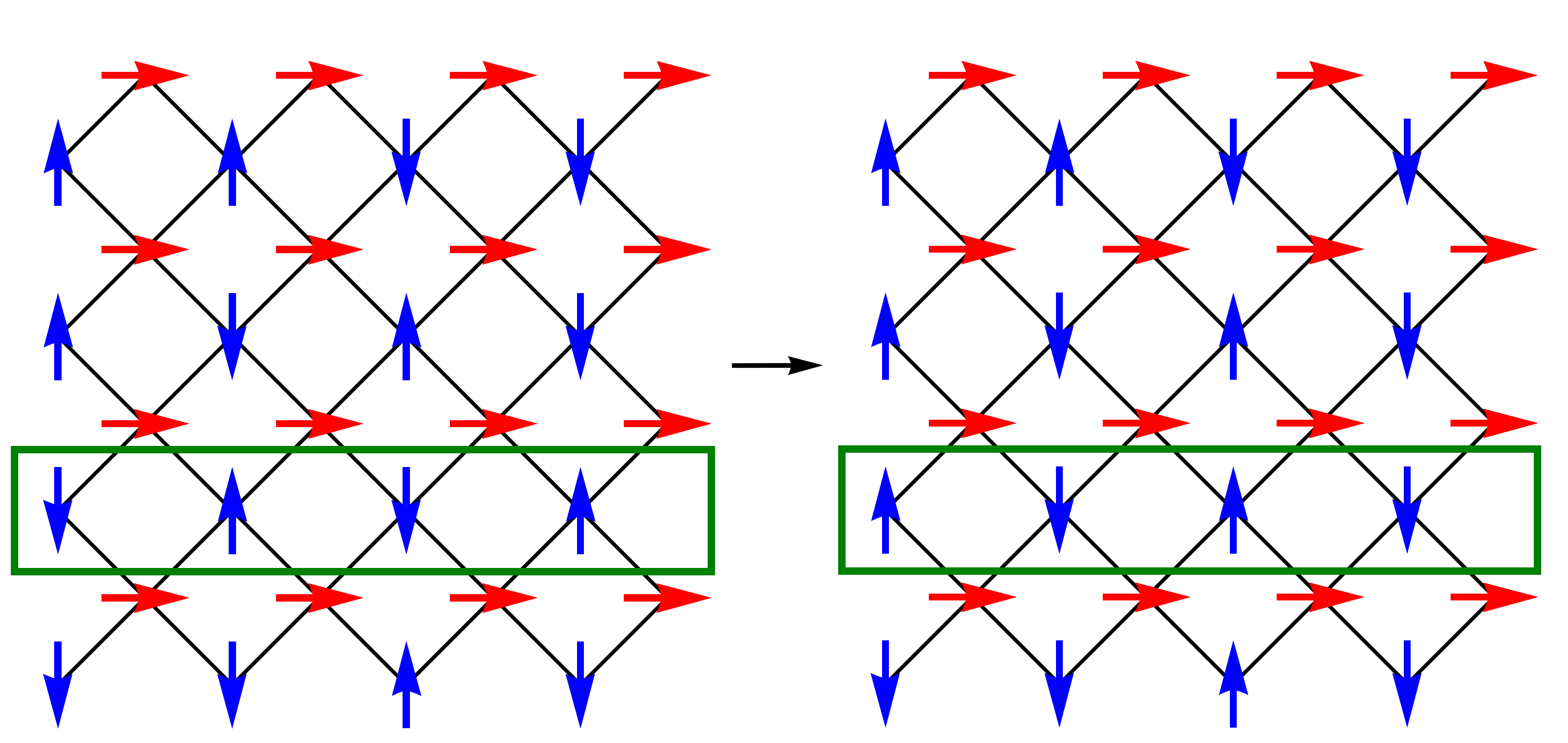}
	\caption{Schematic diagrams of the free-line update. The green line is the region selected to flip the $\tau$ spin along a straight line. Since such a flipping do not change the weight in any given configuration in the 2D cluster model perturbed by $h_x$, its acceptance probability is given $P_{fl}=0.5$ with the heat-bath algorithm.}
	\label{fig:fig_schematic_free_line_update}
\end{figure}

However, such a particular boundary condition at the boundary between $|\Omega\rangle$ and $|\Psi(\alpha)\rangle$ makes the configuration space $\langle\Omega|\Psi(\alpha)\rangle$ become more glassy. As a result, the sampling process in the projector QMC simulation is easy to be stranded in a local minimum configuration. To improve the sampling efficiency, coming out of the subsystem symmetry nature of the 2D cluster model, we introduce a spin update process that sweeping each row and column, and flipping all $\tau^{z}$ along with this row (or column) with probability $P_{fl}=0.5$ (see the green rectangle in Fig.\ref{fig:fig_schematic_free_line_update} for instance). Since flipping all $\tau^{z}$ along $x$- or $y$-axies would not change the sampling weight for the 2D cluster model perturbed by $h_{x}$, the acceptance probability of such a flipping process is $0.5$ according to the heat bath method.

\begin{figure}[htp!]
	\centering
	\includegraphics[width=\columnwidth]{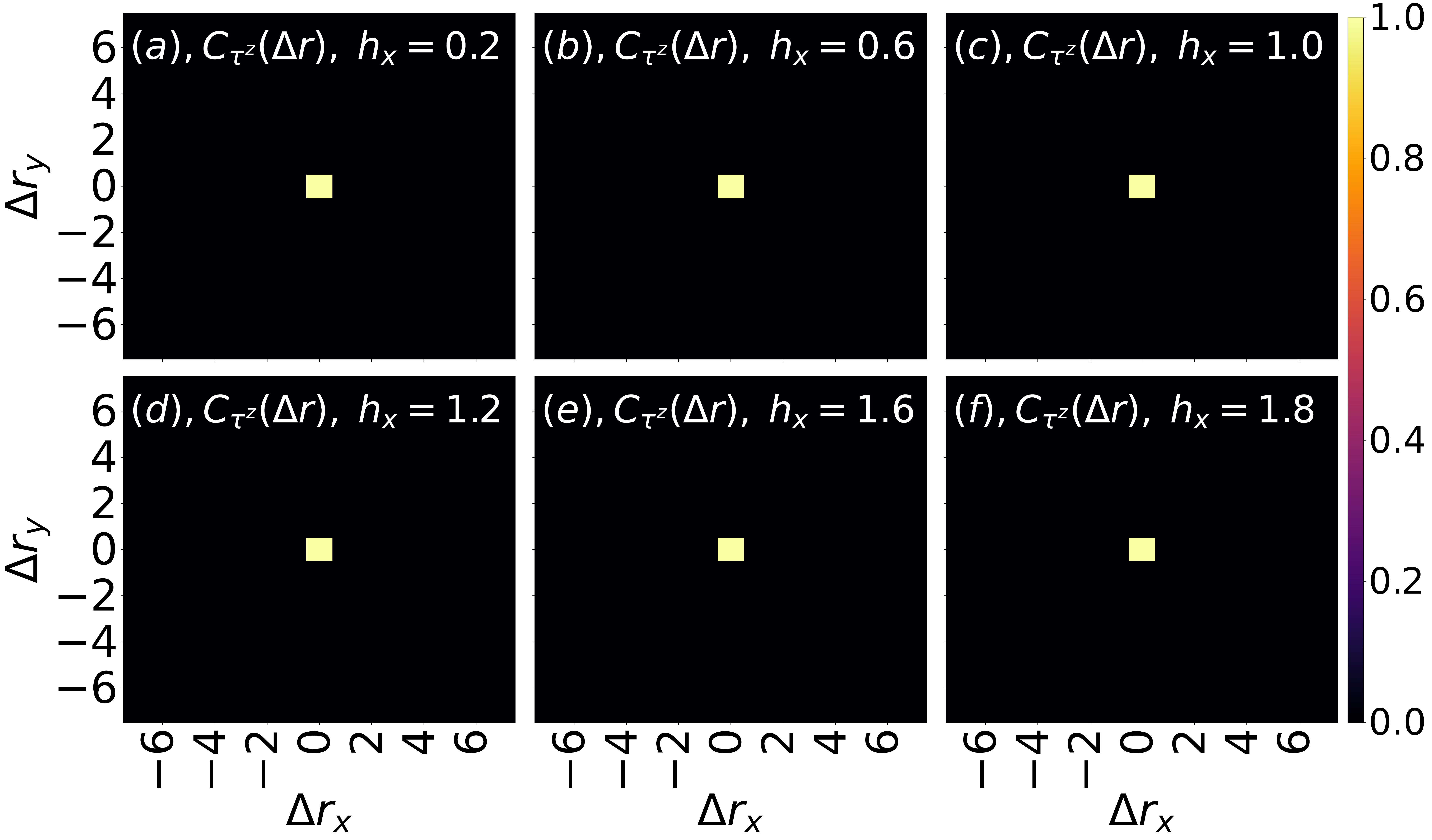}
	\caption{The real-space strange correlator $C^{}_{\tau^{z}}(\Delta r)$ in the strong SSPT model. Panel (a-c) are in the strong SSPT phase with $h_{x}$ changing from $0.2$, to $0.6$, and finally to $1.0$, while the others stand for the trivial paramagnetic phase with $h_{x}$ varying from $1.2$ to $1.8$.}
	\label{fig:fig_real_C_t}
\end{figure}

Beside the strange correlators mentioned in the main part, we also measure the following strange correlators. First, $C_{\tau^{z}}(\Delta r)$, which is
\begin{equation}
	\begin{aligned}
		\phi(i) &=\tau^{z}_{i}, \\
		C_{\tau^{z}}(\Delta r)&=\frac{\langle\Omega|\tau^{z}_{i+\Delta r} \tau^{z}_{i}(-H)^{2n}|\Psi(0)\rangle}{\langle\Omega|(-H)^{2n}|\Psi(0)\rangle}.
	\end{aligned}
	\label{eq:apeq17}
\end{equation}
Fig.\ref{fig:fig_real_C_t} tells the real space dependence of $C_{\tau^{z}}(\Delta r)$, which is no correlation in all direction. Also, it is independent of $h_{x}$.

\begin{figure}[htp!]
	\centering
	\includegraphics[width=0.95\columnwidth]{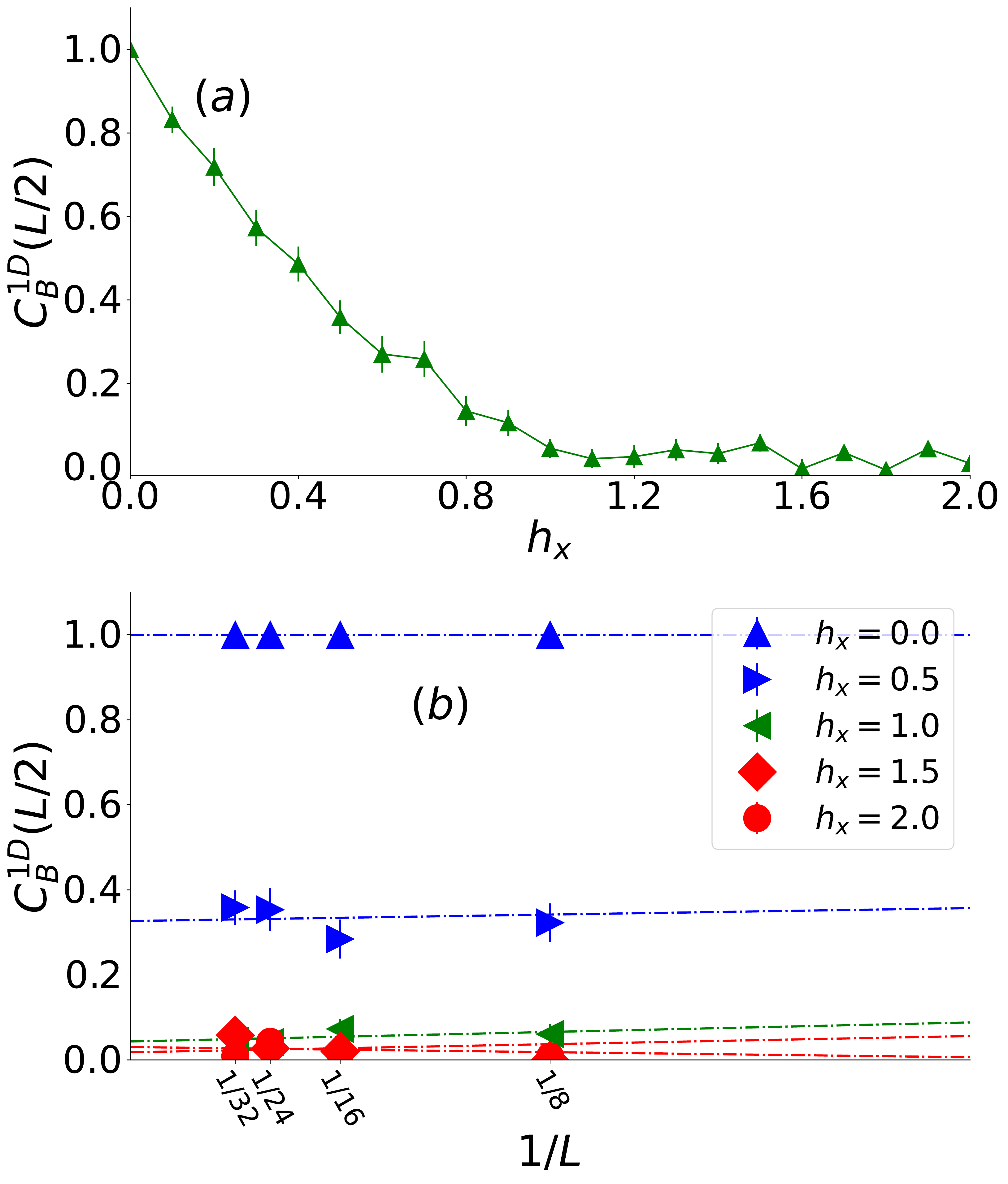}
	\caption{The strange order parameter $C^{1D}_{B}(L/2)$. Panel (a) are $C^{1D}_{B}(L/2)$ plotted along with $h_{x}$ increasing. Panel (b) is the finite-size analyzing of $C^{1D}_{B}(L/2)$, in which $C^{1D}_{B}(L/2)$ is fitting by $C^{1D}_{B}(L/2)(\infty)+c/L$.
	}
	\label{fig:fig_strangr_op_C_B_1D}
\end{figure}

In the 1D cluster model, we also have measured $C^{1D}_{B}$ that
\begin{equation}
	\begin{aligned}
		\phi(i) &=B^{1D}_{i}, \\
		C^{1D}_{B}(\Delta r)&=\frac{\langle\Omega|B^{1D}_{i+\Delta r} B^{1D}_{i}(-H)^{2n}|\Psi(0)\rangle}{\langle\Omega|(-H)^{2n}|\Psi(0)\rangle}.
	\end{aligned}
	\label{eq:apeq18}
\end{equation}
Fig.~\ref{fig:fig_strangr_op_C_B_1D}(a) describe the strange correlator $C_{B}^{1D}(L/2)$ as a function $h_x$ for a $L=32$ system and such ``strange" order parameter also vanishes at the critical point $h^{1D}_{x,c}=1$. Fig.~\ref{fig:fig_strangr_op_C_B_1D} (b) is the extrapolation of $C_{B}^{1D}(L) = C_{B}^{1D}(L=\infty)+a/L$. In the SPT phase ($h_{x}=0.0, 0.5$), $C^{1D}_{\tau^z}(\infty)$ is finite. At the quantum critical point $h_{x,c}$ and inside the paramagnetic phase ($h_{x}=1, 1.5,2.0$), $C^{1D}_{B}(L=\infty)$ tends to $0$, consistent with the phase diagram and our bulk data in Fig.~\ref{fig:fig_normal_result_1D_hz}.

As an order parameter, both $C_{\tau^{z}}^{1D}$ and $C_{B}^{1D}$ can be both applied to tell the SPT phase. However, since $C_{\tau^{z}}^{1D}$ is a simply two spin correlation, we preform the $C_{\tau^{z}}^{1D}$ in the main part but $C_{B}^{1D}$ here.

\section{Strange correlators at the exactly solvable point}
\label{sec:exact_calculation}

In this Appendix, we demonstrate how to analytically obtain the strange correlators at the exactly solvable points of 2D and 1D cluster models as in Table~\ref{table:sc_strong} and Tabel.~\ref{table:sc_weak}. As there is no risk of introducing ambiguity, here we use $|\Psi\rangle$ to refer to the ground state for both 2D and 1D cluster models. And for convenience, in this section we set $C_{\phi}(\Delta r) = C_{\phi}(i,j)$ (and $C^{1D}_{\phi}(\Delta r) = C^{1D}_{\phi}(i,j)$ for 1D cluster model), where $\Delta r=i-j$ is the displacement between $i$ and $j$, and $i$ ($j$) is the site where $\phi_i$ ($\phi_j$) acts. Here we noticed that the results of $\phi=\tau^z,\sigma^z$ case in 1D cluster model has already been analytically obtained in Ref.~\cite{Ellison2021symmetryprotected}.

In the following computation, $i\neq j$ is always assumed. When $i=j$, we can obviously obtain $C_{\phi}(i,j)=1$.  At first, we consider the 2D cluster model:
\begin{itemize}
	\item $\phi=B,A$: In this case, the ground state $|\Psi\rangle$ satisfies $\phi_{i}\phi_{j}|\Psi\rangle=|\Psi\rangle$ according to the definition of the ground state (see Appendix~\ref{appendix_gs})
	, so we can obtain that $\langle\Omega|\phi_{i}\phi_{j}|\Psi\rangle=\langle\Omega|\Psi\rangle$,
	thus $C_{\phi}(i,j)=1$.
	\item $\phi=\tau^{x},\sigma^{x}$: In this case, the trivial state $\langle\Omega|$
	satisfies $\langle\Omega|\phi_{i}\phi_{j}=\langle\Omega|$, so we
	can obtain that $\langle\Omega|\phi_{i}\phi_{j}|\Psi\rangle=\langle\Omega|\Psi\rangle$,
	thus $C_{\phi}(i,j)=1$.
	\item $\phi=\tau^{z},\sigma^{z}$: Without loss of generality, we set $\phi=\tau^{z}$.
	As discussed in Appendix~\ref{appendix_gs},
	$|\Psi\rangle$ can be recognized a equal-weight superposition of Ising configurations of $\{\sigma^{z}\}$ with $\tau^{x}=-1$ decorated
	at the corners of domain walls between
	$\sigma^{z}$'s with opposite values. As $\langle\Omega|\phi_{i}\phi_{j}$
	is a state with $\tau^{x}=-1$ for exactly two $\tau$ spins and $\tau^{x}=1$ for the others, and it is impossible to
	find an Ising configuration with exactly two corners of domain walls in 2D, $\langle\Omega|\phi_{i}\phi_{j}$
	can only have zero overlap with an arbitrary configuration from $|\Psi\rangle$.
	Thus $C_{\phi}(i,j)=0$. Similarly, we can obtain $C_{\phi}(i,j)=0$ for $\phi=\sigma^{z}$.
	\item $\phi=\tau_{i}^{z}\tau_{i+\hat{y}}^{z},\sigma_{i}^{z}\sigma_{i+\hat{y}}^{z}$:
	Without loss of generality, we set $\phi=\tau_{i}^{z}\tau_{i+\hat{y}}^{z}$.
	At first, when $i$ and $j$ are located on the same straight line
	exactly along $x$ direction, we can notice that $\langle\Omega|\phi_{i}\phi_{j}=\langle\Omega|\prod_{k\in S}B_{k}$,
	where $S$ is a straight string connecting $i$ and $j-\hat{x}$ (here we assume the $x$-coordinate of $j$ is larger than of $i$), because the $\sigma^x$ operators in $B$ act on $\langle \Omega|$ trivially. So $\langle\Omega|\phi_{i}\phi_{j}|\Psi\rangle=\langle\Omega|\prod_{k\in S}B_{k}|\Psi\rangle=\langle\Omega|\Psi\rangle$, where the second equality is according to the definition of the ground state $|\Psi\rangle$ (see Appendix~\ref{appendix_gs}),
	thus $C_{\phi}(i,j)=1$. If $i$ and $j$ do not satisfy the above condition,
	then following the same logic as in the $\phi=\tau^{z},\sigma^{z}$
	case, $\langle\Omega|\phi_{i}\phi_{j}$is a state with $\tau^{x}=-1$ for exactly four $\tau$ spins and $\tau^{x}=1$ for the others, however, such four sites with $\tau^{x}=-1$ cannot
	form the corners of domain walls of any Ising configurations, so $\langle\Omega|\phi_{i}\phi_{j}$
	can only have zero overlap with an arbitrary configuration from $|\Psi\rangle$.
	Thus $C_{\phi}(i,j)=0$ (this result can also be obtained based on the behavior of $D_iD_j$ under symmetry transformations, see Sec.~\ref{sec:results}). In conclusion, for $i$ and $j$ on the same straight
	line along $x$ direction, $C_{\phi}(i,j)=1$, otherwise $C_{\phi}(i,j)=0$. The same results can be obtained for $\phi=\sigma^z_i \sigma^z_{i+\hat{y}}$.
\end{itemize}

Then, for the 1D cluster model case, we have:
\begin{itemize}
	\item $\phi=B^{1D},A^{1D}$: In this case, the ground state $|\Psi\rangle$ satisfies $\phi_{i}\phi_{j}|\Psi\rangle=|\Psi\rangle$ according to the definition of the ground state (see Sec.~\ref{sec:1d_case})
	, so we can obtain that $\langle\Omega|\phi_{i}\phi_{j}|\Psi\rangle=\langle\Omega|\Psi\rangle$,
	thus $C^{1D}_{\phi}(i,j)=1$.
	\item $\phi=\tau^{x},\sigma^{x}$: In this case, the trivial state $\langle\Omega|$
	satisfies $\langle\Omega|\phi_{i}\phi_{j}=\langle\Omega|$, so we
	can obtain that $\langle\Omega|\phi_{i}\phi_{j}|\Psi\rangle=\langle\Omega|\Psi\rangle$,
	thus $C^{1D}_{\phi}(i,j)=1$.
	\item $\phi=\tau^{z},\sigma^{z}$: Without loss of generality, we set $\phi=\tau^{z}$.
	We can notice that $\langle\Omega|\phi_{i}\phi_{j}=\langle\Omega|\prod_{k\in S}B^{1D}_{k}$,
	where $S$ is a string composed of $\sigma$ spins connecting $i$ and $j-1$ (here we set a unit cell to be composed of a $\tau$ spin at the left and a $\sigma$ spin at the right, and $j>i$ is assumed), because the $\sigma^x$ operators in $B^{1D}$ act on $\langle \Omega|$ trivially. So $\langle\Omega|\phi_{i}\phi_{j}|\Psi\rangle=\langle\Omega|\prod_{k\in S}B^{1D}_{k}|\Psi\rangle=\langle\Omega|\Psi\rangle$, where the second equality is according to the definition of the ground state $|\Psi\rangle$ (see Sec.~\ref{sec:1d_case}),
	thus $C^{1D}_{\phi}(i,j)=1$. Similarly, we can obtain $C^{1D}_{\phi}(i,j)=1$ for $\phi=\sigma^{z}$.
\end{itemize}

\renewcommand{\theequation}{S\arabic{equation}}
\renewcommand{\thefigure}{S\arabic{figure}}

\newpage

\end{document}